\documentclass[fleqn,usenatbib]{mnras}

\usepackage{newtxtext,newtxmath}
\usepackage[T1]{fontenc}
\usepackage{ae,aecompl}

\usepackage{graphicx}	% Including figure files
\usepackage{amsmath}	% Advanced maths commands
\usepackage{amssymb}	% Extra maths symbols

\def\bs#1{\mbox{\boldmath $#1$}}
\def\rref#1{(\ref{#1})}

\newcommand{\om}{\Omega_{\rm m0}}

\newcommand{\Msun}{{\rm M}_{\odot}}

\newcommand{\mean}[1]{\langle #1 \rangle}
\newcommand{\mizuki}{{\texttt{Mizuki}}}
\newcommand{\demp}{{\texttt{DEmP}}}

\title[Effect of magnification on type Ia supernova]%
{Effect of lensing magnification on type Ia supernova cosmology}

\author[H. Sakakibara et al.]
{Hinako Sakakibara,$^{1}$\thanks{E-mail: sakakibara.hinako@f.mbox.nagoya-u.ac.jp}
Atsushi  J. Nishizawa,$^{1,2}$\thanks{E-mail: atsushi.nishizawa@iar.nagoya-u.ac.jp}
Masamune Oguri,$^{3,4,5}$
\newauthor
Masayuki Tanaka,$^{6}$
Bau-Ching Hsieh$^{7}$ and
Kenneth C. Wong$^{5}$
\\
$^{1}$ Department of Physics, Nagoya University, Furocho, Chikusa, Nagoya 464-8602, Aichi, Japan\\
$^{2}$ Institute for Advanced Research, Nagoya University, Furocho, Chikusa, Nagoya 464-8602, Aichi, Japan\\
$^{3}$ Research Center for the Early Universe, University of Tokyo, Tokyo 113-0033, Japan\\
$^{4}$ Department of Physics, University of Tokyo, Tokyo 113-0033, Japan \\
$^{5}$ Kavli Institute for the Physics and Mathematics of the Universe (Kavli IPMU, WPI), University
of Tokyo, Chiba 277-8582, Japan\\
$^{6}$ National Astronomical Observatory Japan, 2-21-1 Osawa, Mitaka, Tokyo 181-8588, Japan\\
$^{7}$ Academia Sinica Institute of Astronomy and Astrophysics, P.O. Box 23-141, Taipei 10617, Taiwan
}

\date{Accepted XXX. Received YYY; in original form ZZZ}

\pubyear{2018}

\begin{document}
\label{firstpage}
\pagerange{\pageref{firstpage}--\pageref{lastpage}}
\maketitle

\begin{abstract}
Effect of gravitational magnification on the measurement of distance modulus of type Ia supernovae is presented.
We investigate a correlation between magnification and Hubble residual to explore how the magnification affects the estimation of cosmological parameters.
We estimate magnification of type Ia supernovae in two distinct methods:
one is based on convergence mass reconstruction under the weak lensing limit and the other is based on the direct measurement from galaxies distribution. Both magnification measurements are measured from Subaru Hyper Suprime-Cam survey catalogue.
For both measurements, we find no significant correlation between Hubble residual and magnification.
Furthermore, we correct for the apparent supernovae fluxes obtained by Supernova Legacy Survey 3-year sample using direct measurement of the magnification.
We find $\om = 0.287 ^{+0.104} _{-0.085}$ and $w = -1.161 ^{+0.595}_{-0.358}$ 
for supernovae samples corrected for lensing magnification 
when we use photometric redshift catalogue of {\mizuki}, while
$\om = 0.253 ^{+0.113} _{-0.087}$ and $w = -1.078 ^{+0.498} _{-0.297}$ 
for {\demp} photo-z catalogue.
Therefore, we conclude that the effect of magnification on the supernova cosmology is negligibly small for the current surveys; however, it has to be considered for the future supernova survey like LSST.
\end{abstract}


%%%%%%%%%%%%%%%%%%%%%%%%%%%%%%%%%%%%%%%%%%%%%%%%%%

%%%%%%%%%%%%%%%%% BODY OF PAPER %%%%%%%%%%%%%%%%%%

%%%%%%%%%%%%%%%%%%%%%%%%%%%%%%%%%%%%%%%%%%%%%%%%%%%%%%%%%%%%%%%%%%%%%
\section{introduction}
\label{sec:introduction}
%%%%%%%%%%%%%%%%%%%%%%%%%%%%%%%%%%%%%%%%%%%%%%%%%%%%%%%%%%%%%%%%%%%%%
Type Ia supernova (SN) is a useful tool to probe cosmological model. The absolute magnitude of SN at the cosmological distance is empirically well calibrated with the local SN and due to the brightness, it can be observed up to high redshift $z\gtrsim 1$. 
Several SNe surveys over wide range of redshift are carried out 
in this two decades for cosmological study
\citep{Riess+:1999,Filippenko+:2001,Astier+:2006,Frieman+:2008,
Dawson+:2009,Grogin+:2011}.
Since late 1990s, distant SNe Ia surveys have suggested that the Universe is 
accelerating expanding
\citep{Garnavich+:1998,Riess+:1998,Perlmutter+:1999}.
The accelerating expansion of the Universe requires the existence of dark energy within the context of general relativity or possible extension of the theory of gravity.
Lately, other cosmological probes
such as cosmic microwave background 
\citep[CMB;][]{Komatsu+:2011,Planck2015:cosmology} and baryon acoustic oscillation 
\citep[BAO;][]{Percival+:2010,Alam+:2017} also 
prefer the cosmological model consistent with what obtained by the SNe observations. 

With the SNe Ia, we can constrain the cosmological models by use of distance modulus over their redshifts by comparing the difference between apparent and absolute magnitude to the theoretical prediction which simply can be described by the luminosity distance
\citep{Astier+:2006,Kessler+:2009,Guy+:2010,Suzuki+:2012,Ganeshalingam+:2013,
Rest+:2014}.
When we look at the distance modulus around the best predicted curve, it is prominently observed that there is a large scatter around the prediction. Apart from the statistical fluctuation, the scatters originate both from intrinsic diversity of the SNe, and the flux magnification due to the gravitational lensing by the foreground mass distribution.
The SNe fluxes are amplified when the local matter density along the line of sight of SN is higher than average, while they are diminished at lower density regions.
Several attentions have been paid to the effect of gravitational lensing on the measurement of the distance modulus of SNe Ia.
\cite{Frieman:1996} 
found
that the dispersion of apparent magnitude 
due to weak lensing by large scale structure is 
$\sigma \sim 0.04\om^{1/2}$ at source redshift $z=0.5$ 
in the flat $\Lambda$CDM model.
\cite{HamanaFutamase:2000} considered the lensing dispersion in peak magnitude 
of SNe 
and found that
the dispersion $\sigma \sim 0.057$ at redshift $z=1$.
\cite{Gunnarsson+:2006} made mock galaxy data to investigate the lensing dispersion 
and showed that a correction of lensing effect can reduce the 
lensing dispersion from about 7\% to 3\% for a source at $z=1.5$.
\cite{HadaFutamase:2016} predicted that the lensing 
dispersion of SNe Ia data sets is about 0.03 mag at $z=1$ by using galaxies 
with virial mass $M_{\rm vir} > 10^{11}M_{\odot}$.

The expected effect of gravitational lensing on the SN flux has been studied by using numerical simulations.
\cite{Wambsganss+:1997} found that the gravitational lensing effect causes 
the magnitude dispersion of source objects of 0.04 mag at redshift $z=1$.
\cite{Jonsson+:2009} have found that with the mock simulation assuming SNLS-like survey can reduce the scatter on the SNe fluxes by 4\% when the intrinsic scatter is 0.13 mag and that the errors on $\om$ and $w$ can be reduced by $4-6\%$.
The probability distribution functions (PDFs) of lensing magnification 
are also investigated by simulations.
\cite{Wang:1999} proposed a fitting formula of PDF for the magnification $p(\mu_{\rm lens}|z_{\rm SN})$ as the function of $\om, \Omega_{\Lambda 0}$ and the parameter 
$\tilde{\alpha}$ representing the inhomogeneous density distribution.
\cite{Takahashi+:2011} studied PDFs of convergence, shear and 
magnification by performing N-body ray-tracing simulations to 
find the relation between mean and variance of convergence and 
found the analytic formulae of PDFs well fit the simulated ones.

Also there have been several attempts for correcting the magnification applied to the actual observed data.
\cite{Jonsson+:2007} used 26 SNe in GOODS fields to study the correlation 
between lensing magnification and residual magnitude of SNe.
They found that the correlation coefficient between them is $r=0.29$ with rather weak significance of $90\%$, which is consistent with what expected given the small number of samples.
\cite{Kronborg+:2010} used deep images of CFHTLS with SNLS SNe and found a lensing signal at $2.3\sigma$ significance.
\cite{Smith+:2014} calculated convergence by using SDSS-II galaxy sample 
for each SNe Ia from SDSS-II and BOSS surveys under the weak lensing 
approximation and found a correlation between convergence 
and Hubble residual 
at $1.7\sigma$ significance.

%It has not been studied the effect of gravitational lensing along line of sight of SN 
%on constraint of cosmological parameters using high-redshift SN data sets.
%For future survey which can detect large number of SNe Ia at $z > 1$, 
%the gravitational lensing effect has an important role to constrain 
%cosmological parameters precisely.
%The aim of our study is to investigate the lensing effect on estimation of 
%cosmological parameters using SNLS3 SNe Ia data sets and HSC galaxy data sets .

In this paper, we investigate magnification of SNe Ia 
fluxes using HSC galaxy photometric catalog. We estimate magnification by two distinct methods: one measures convergence by mass 
reconstruction using galaxy shape catalog under the weak lensing approximation.
The other method measures magnification directly from galaxy distribution assuming that galaxy resides dark matter halo with an NFW profile without assuming the weak lensing regime.
We also investigate the impact of magnification effect on 
the cosmological parameter estimation by comparing the results with and without the magnification correction to the distance modulus.

The paper is organized as follows. In section \ref{sec:data}, we describe 
the SNe Ia and galaxy data sets we use. 
In section \ref{sec:sim}, we describe our simulation used for
evaluating the expected magnification.
In section \ref{sec:dm}, we overview
the measurement of the distance modulus. In section \ref{sec:est_mu}, we
revisit the 
measurement of the magnification based on the shape catalog 
and then introduce
our new estimator which better expresses the magnification.
In section \ref{sec:result}, we present the results of magnification by 
two different estimators and the effect of magnification on the parameter estimation.
In section \ref{sec:summary}, we give a conclusion and summary.

%%%%%%%%%%%%%%%%%%%%%%%%%%%%%%%%%%%%%%%%%%%%%%%%%%%%%%%%%%%%%%%%%%%%%
\section{Data sets}
\label{sec:data}
%%%%%%%%%%%%%%%%%%%%%%%%%%%%%%%%%%%%%%%%%%%%%%%%%%%%%%%%%%%%%%%%%%%%%

%----------------------------------------------------------------------------
\subsection{Supernova Legacy Survey 3-year sample}
\label{ssec:SNLS}
%----------------------------------------------------------------------------
We use Supernova Legacy Survey (SNLS) 3-year data products for SNe Ia analysis
\citep{Guy+:2010}.
SNLS program is carried out from 2003 to 2008 to detect high-redshift 
SNe Ia which are then used for constraining the cosmological model such as dark energy.
SNLS consists of two distinct observations:
one is photometric cadence survey to measure 
the light curves of the SNe
and the another is spectroscopic survey to 
confirm the redshift and the spectral type of the detected transient objects.
SNLS uses the deep images of MegaCam on Canada-France-Hawaii Legacy Survey (CFHTLS), which cover four distinct patches of the sky. These patches are named \texttt{D1} to \texttt{D4} and roughly one square degree area for each.
The images are taken in four broad band filters to obtain the color information of the SN.
The SNe spectra are taken by Very Large Telescope (VLT), 
Gemini-North and South and Keck telescopes.

SNLS 3-year data sets are used to estimate cosmological parameters.
From SNLS3 only, 
\cite{Guy+:2010} %found 
find
$\om = 0.211 \pm 0.034 ({\rm stat.}) \pm 0.069 ({\rm sys.})$
and \cite{Conley+:2011} %found 
extend the analysis to time varying dark energy model to find
the 
equation-of-state parameter
$w = -0.91 ^{+0.16}_{-0.20} ({\rm stat.}) ^{+0.07}_{-0.14}({\rm sys.})$.
A joint analysis of SNLS3 with BAO from Sloan Digital Sky Survey (SDSS) and CMB from 7 year Wilkinson Microwave Anisotropy Probe (WMAP) yields
$\om = 0.269 \pm 0.015$ and
$w = −1.061 ^{+0.069}_{-0.068}$ for the flat Universe model \citep{Sullivan+:2011}.
SNLS3 data sets are also used to constrain SNe Ia progenitors \citep{Bianco+:2011} 
and to improve the accuracy of photometric calibration \citep{Betoule+:2013}.

\cite{Guy+:2010} use two empirical methods for modelling SN Ia light curve: 
SALT2 \citep{Guy+:2007}, and SiFTO \citep{Conley+:2008}.
The main difference between two is in the way of modeling the color correction term $C$, defined in equation (\ref{eq:observed_dm}). The SALT2 uses a single color to constrain $C$ from the light curve fitting while SiFTO uses 5 filters.

Following \cite{Guy+:2010}, we make a clean sample of SNe by removing SNe having poorly constrained light curve, extreme property among the diversity of type Ia SNe, peak color is significantly affected by dust extinction in the host galaxy or the number of colors obtained is significantly smaller than what has been scheduled.
We also exclude outliers in the Hubble residual and finally we obtain 231 SNe.

%----------------------------------------------------------------------------
\subsection{Hyper Suprime-Cam}
\label{ssec:HSC}
%----------------------------------------------------------------------------
Hyper Suprime-Cam (HSC), 
is the wide field optical imaging camera installed on the prime focus of the Subaru Telescope \citep{Miyazaki+:2018, Komiyama+:2018, Kawanomoto+:2018, Furusawa+:2018}. The field of view of the camera is 1.77 square degrees and the pixel size is 0.17 arcsecs. With this gigantic camera and the good quality of seeing and transparency, HSC collects the precise shape and photometry of galaxies.
The HSC survey is a Subaru Strategic Program (SSP) which consists of three different layers, wide (1400 [deg$^2$]), deep (27.5 [deg$^2$]) and ultra-deep (3.5 [deg$^2$]), with five broad-band filters, g,r,i,z and y. Deep and ultra-deep layers have additional four narrow-band filters \citep{AiharaArimoto+:2018}.

In this paper, we use photometric redshift catalogs for S17A release from deep and ultra-deep layers and S16A galaxy shape catalog from wide layers overlapped with deep and ultra-deep layers \citep{Mandelbaum+:2018}.
The shape of the galaxies are measured based on the re-Gaussianization method \citep{HirataSeljak2003}, applied to the coadd of $i$-band images in the full-depth-full-color (FDFC) regions where all the broad-band data reaches to the target depthes (26 PSF magnitude in $i$-band). Interested readers should refer to \cite{Mandelbaum+:2008} for more detail about the shape catalog.

We use two different photometric redshift catalogs: one from template fitting, {\mizuki} \citep{Tanaka:2015} and the other based on the empirical method, {\demp} \citep{HsiehYee:2014}.
Both codes are calibrated with the publicly available spectroscopic redshifts, grism/prism redshift and high quality photometric redshift in COSMOS \citep{Laigle2016}.
{\mizuki} derives photometric redshift and stellar mass simultaneously by fitting the 5 band photometry to the expected galaxy templates, while the {\demp}  derives redshift by looking for 40 nearest counterpart of the spectroscopic redshift in color-magnitude multi-dimensional space. {\demp} also finds the stellar mass by the same way but in the COSMOS high precision photo-z catalog in color-magnitude-redshift multi-dimensional space.
The photometric redshift accuracy and methodology are summarized in \cite{Tanaka+:2018}.

The HSC and SNLS fields are partially overlapped at \texttt{D1, D2} and \texttt{D3} fields, which contains 158 sample selected
SNe. We further remove 5 SNe located near the very bright stars, with the separation closer than 0.8 arcmins, because the photometry of galaxies in the vicinity of bright star can be contaminated by the star and brings large systematic errors on the photometric redshifts. The example of the SN near the bright star is shown in the left panel of Figure \ref{fig:usegalaxy}. 
Eventually, the number of SNe we use in our analysis is 153.

%%%%%%%%%%%%%%%%%%%%%%%%%%%%%%%%%%%%%%%%%%%%%%%%%%%%%%%%%%%%%%%%%%%%%
\section{Simulation}
\label{sec:sim}
%%%%%%%%%%%%%%%%%%%%%%%%%%%%%%%%%%%%%%%%%%%%%%%%%%%%%%%%%%%%%%%%%%%%%
In this section, we describe the expectation of the amount of magnification due to the large-scale structure both from numerical simulation and the standard $\Lambda$CDM prediction.

%----------------------------------------------------------------------------
\subsection{Specifications of the suite of simulation}
\label{ssec:sim_data}
%----------------------------------------------------------------------------
We use the all-sky multiple lens plane ray-trace simulation data sets 
\citep{Takahashi+:2017}.
They first make high-resolution N-body simulations with 14 different boxes at light cone output placed around the observer. The box sizes are from 450 to 6300 $h^{-1}$Mpc with interval of $450\, h^{-1}$Mpc. Redshift of each output corresponds to the radial distance from the observer at centre.
The number of particles included in each box is fixed to $2048^3$ so that the mass resolution is higher at lower redshift.
Each cubic box is divided into three spherical shells separating $\Delta r=150 h^{-1}$Mpc and the particle positions are projected onto the shells. The entire sky is segmentalized by the \texttt{Healpix} pixels into $12\times N_{\rm side}^2$ equal area pixels, and all the particle positions are replaced by the central position of the nearest pixel. The projected density can be used to compute the deflection angle and then complete a multiple lens approximated ray-tracing simulation. We use $N_{\rm side}=4096$, which corresponds to the pixel of 0.86 arcmins on a side up to redshift $z=1.033$.

%----------------------------------------------------------------------------
\subsection{Expectation from the simulation and $\Lambda$CDM model}
\label{ssec:sim_theory}
%----------------------------------------------------------------------------
In this subsection, we estimate analytic $\delta\mu_{\rm lens} \equiv \mu_{\rm lens} - 1$ 
from $\Lambda$CDM model and from N-body simulation.
Here we assume the flat $\Lambda {\rm CDM}$ Universe.
The matter perturbation along the line of sight makes the image of 
the source object distorted.
The convergence at given sky position $\bs{\theta}$ along the line of sight is approximated by \citep{BartelmannSchneider:2001},
\begin{equation}
	\label{eq:convergence}
	\kappa(\bs{\theta}) =
    \kappa_0
	\int_0^{\chi_s} \! {\rm d}\chi 
	\frac{\chi (\chi_s -\chi)}{\chi_s} \frac{\delta(\chi \bs{\theta}, \ \chi)}{a(\chi)},
\end{equation}
where $\chi_s$ is the comoving distance from us to the source object,
$\delta(\chi\bs{\theta}, \ \chi)$ is three dimensional overdensity of matter distribution and $\kappa_0=3H_0^2\om/2c^2$.
The power spectrum of the convergence is obtained by using 
the Limber's equation in the Fourier space \citep{Limber:1954,Kaiser:1992},
\begin{equation}
	\label{eq:P_kappa}
	P_{\kappa}(l)
	=
	\kappa_0^2
	\int_0^{\chi_s} \! {\rm d}\chi
	\left( 1-\frac{\chi}{\chi_s} \right)^2
	\frac{1}{a^2(\chi)}
	P_{\delta}\left( \frac{l}{\chi}, \  \chi \right),
\end{equation}
where $P_{\delta}$ is the matter power spectrum.

In order to compare the theoretical prediction with the simulation, we 
smooth the convergence field with two dimensional Gaussian window function in the Fourier space,
\begin{equation}
	\label{eq:Fourier_window}
	\tilde{W}_{\Theta}(l)
	=
	\exp \left( -\frac{l^2 \Theta ^2}{2} \right),
\end{equation}
where $\Theta$ is a smoothing scale.
Since we use the simulation with $N_{\rm side}=4096$ whose pixel area is $s = 0.738$ arcmin$^2$, 
we adopt $\Theta = 0.485$ arcmin so that $\Theta$ satisfies $\pi \Theta^2 = s$.
Then the power spectrum of the smoothed convergence $\bar{\kappa}$ is 
rewritten as
\begin{equation}
	\label{eq:smoothed_P_kappa}
	P_{\bar{\kappa}}(l;\Theta)
	=
	\kappa_0^2
	\int_0^{\chi_s} \! {\rm d}\chi
	\left( 1-\frac{\chi}{\chi_s} \right)^2
	\frac{1}{a^2}
	P_{\delta}\left( \frac{l}{\chi}, \  \chi \right)
	\tilde{W}_{\Theta}^2(l).
\end{equation}
Assuming the weak gravitational lens, 
the magnification can be approximated by equation \rref{eq:approx_mulens}.
Therefore, the smoothed $\overline{\delta\mu}_{\rm lens}$ power spectrum 
$P_{\overline{\delta\mu}_{\rm lens}}$ is equal to $4P_{\bar{\kappa}}$.

The corresponding two-point angular correlation function is related to 
$P_{\overline{\delta\mu}_{\rm lens}}$ with
\citep{Peebles:1973},
\begin{equation}
	\label{eq:angular_corr}
	w_{\overline{\delta\mu}_{\rm lens}}(\theta)
	=
	\frac{1}{4\pi}
	\sum_{l=0}^{\infty} (2l+1)P_{\overline{\delta\mu}_{\rm lens}}(l) \mathcal{P}_l(\cos\theta),
\end{equation}
where $\mathcal{P}_l$ is the Legendre polynomial. 
Then the root mean square of $\overline{\delta \mu}_{\rm lens}$ is readily obtained as 
\begin{equation}
	\label{eq:rms_deltamu}
	\overline{\delta\mu}_{\rm lens}^{\rm rms}
	=
	\sqrt{\langle \overline{\delta\mu}_{\rm lens}^2 \rangle}
	=
	\sqrt{w_{\overline{\delta\mu}_{\rm lens}}(0)}.
\end{equation}

\begin{figure}
\includegraphics[width=\linewidth]{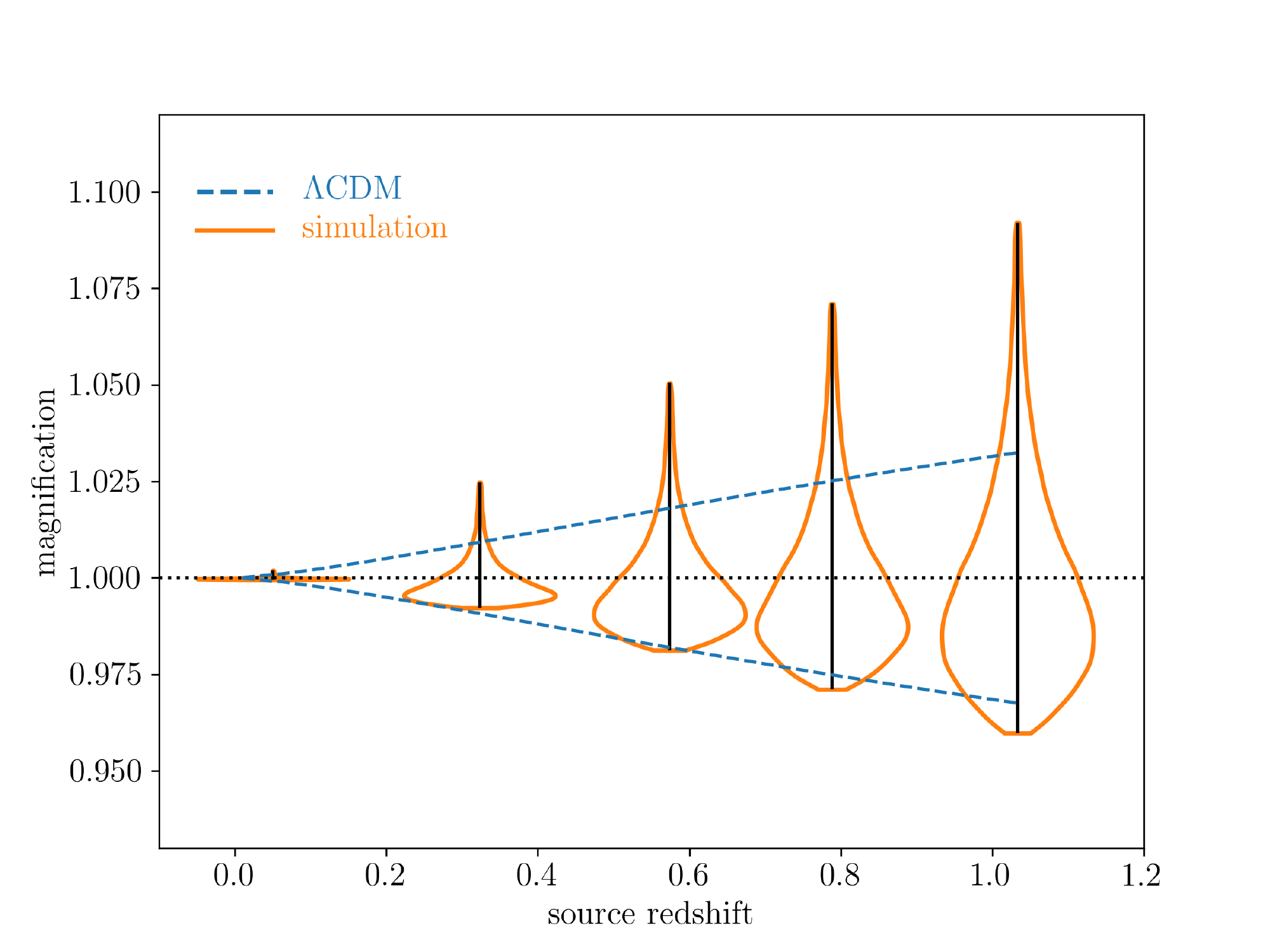}
\caption{
	The relation between source redshift and magnification 
	predicted by $\Lambda$CDM model and simulation.
	The blue dashed lines show $1 \pm \bar{\delta\mu}_{\rm lens}^{\rm rms}$ predicted by $\Lambda$CDM model.
	The orange plots are the probability distributions of magnification obtained by 
	simulation at source redshift $z_s = 0.051, 0.323, 0.574, 0.789$ and $1.033$, truncated at $2\sigma$ limits.}
    \label{fig:LCDM_sim}
\end{figure}

Figure \ref{fig:LCDM_sim} shows the dependence of source redshift on magnification 
predicted by $\Lambda$CDM model and simulation.
At $z_s = 1.033$, $\Lambda$CDM model predicts $\overline{\delta\mu}_{\rm lens}^{\rm rms} = \pm 0.032$, while
from the simulation, we obtained 68\% confidence region as $\pm 0.026$,
which is slightly smaller than that from $\Lambda$CDM prediction.

%%%%%%%%%%%%%%%%%%%%%%%%%%%%%%%%%%%%%%%%%%%%%%%%%%%%%%%%%%%%%%%%%%%%%
\section{Distance modulus for type Ia supernovae}
\label{sec:dm}
%%%%%%%%%%%%%%%%%%%%%%%%%%%%%%%%%%%%%%%%%%%%%%%%%%%%%%%%%%%%%%%%%%%%%
In this section, we overview the methodology of constraining dark
energy model from the distance modulus of type Ia supernovae.

%----------------------------------------------------------------------------
\subsection{distance modulus and dark energy}
\label{ssec:darkenergy}
%----------------------------------------------------------------------------
Distance modulus of SN Ia can be used as the probe of the cosmological model.
The distance modulus is a relation between observed and absolute magnitude which is given by
\begin{equation}
	\label{eq:dm}
	\mu 
	\equiv
	m-M
	=
	5\log_{10}\left( D_L [{\rm pc}] \right)-5,
\end{equation}
where $m$ and $M$ are apparent magnitude and absolute magnitude respectively. 
In the case of type Ia SN, we can readily estimate the absolute magnitude by the method described in \ref{ssec:lightcurve}.
The observed distance modulus can then be compared with the theoretical prediction to constrain the cosmological models.
The luminosity distance in Eq. (\ref{eq:dm}) can be described as
\begin{equation}
	\label{eq:d_lum}
	D_L(z) = (1+z)r(\chi),
\end{equation}
where $r(\chi)$ is the radial coordinate which can be related with comoving distance $\chi$ as
\begin{align}
	\label{eq:d_phys}
	r(\chi) =
	\begin{cases}
		K^{-1/2}\sin (K^{1/2}\chi) & (K > 0) \\
		\chi & (K=0) \\
		(-K)^{-1/2}\sinh [(-K)^{1/2}\chi] & (K < 0)
	\end{cases}
\end{align}
depending on the spatial curvature of the Universe, $K$. The comoving distance is an integral of the inverse of Hubble parameter, 
\begin{align}
	\label{eq:hubble}
	\frac{H^2(z)}{H_0^2}
	=
	&{}\om(1+z)^3+\Omega_{\rm K0}(1+z)^2 \nonumber \\
	&{}+\Omega_{\rm DE0}\exp\left[
    \int_0^z \! 3(1+w
    (z')
    ) {\rm d}\ln (1+z')\right],
\end{align}
where $H_0$ is the current Hubble parameter, 
$\Omega_{\rm m0},\Omega_{\rm K0},$ and $\Omega_{\rm DE0}$ 
are the 
current density parameters of the 
matter, curvature, and dark energy respectively.
In this paper, we consider the 
time-dependent 
dark energy model which can be parametrized by equation of state parameter $w(z)$.

%----------------------------------------------------------------------------
\subsection{light curve of the type Ia supernova}
\label{ssec:lightcurve}
%----------------------------------------------------------------------------
Type Ia SNe are quite accurate standard candles in the Universe so that they can be used to constrain the model of dark energy. Although the binary system of their progenitor evolution is still not clear, SN Ia is a thermonuclear explosion of C+O white dwarf in a binary system \citep[e.g.][]{Maeda+2016}.
As the SNe Ia explosion mechanism and the maximum progenitor mass is limited by the Chandrasekhar mass, the SNe Ia have a uniform luminosity at the peak of the light curve. Although there is some diversity in the intrinsic feature of the SN Ia, the width of the light curve is well correlated with the maximum luminosity and the diversity can be fairly well corrected by the empirical relation. The correction is required to measure the distance accurate enough to constrain the cosmological parameters.
\cite{Phillips:1993} investigated the relation between decline rate of SN Ia 
light curve and peak luminosity, and found that the light curve of brighter SN 
declines slower than the fainter one.
\cite{Riess+:1995} applied 
the relation to distance modulus
and obtained smaller dispersion 
of Hubble residual by a factor 2.4 than the dispersion without correction.
In addition to 
this shape-luminosity relation, 
the relation between luminosity and color is also used to correct the 
absolute magnitude.
\cite{Wang+:2005} found that the peak luminosity is linearly correlated with $B-V$ color and that the correction can reduce the dispersion to 0.18 mag in $V$ band.

From the light curve fitting by \texttt{SiFTO}, rest-frame magnitude $m_B^*$, shape parameter 
$\Gamma$ and color parameter $C$ are obtained. 
Then the corrected distance modulus of SN Ia is defined as
\citep{Guy+:2010},
\begin{equation}
	\label{eq:observed_dm}
	\mu^{\rm obs} = m_B^* - M + \alpha \Gamma - \beta C.
\end{equation}
The measurement uncertainty 
is $\sigma^2(\mu^{\rm obs}) = {\bf V}^{\rm T} {\bf Cov}({\bf X}_s){\bf V}$, where
\begin{equation}
	\label{eq:dm_params}
    {\bf X}_s^{\rm T}
    =
    \left(m_{B,s}^*, \Gamma_s, C_s \right), \ \ 
   	{\bf V}^{\rm T}
    =\left(
	1, \alpha, -\beta \right)
\end{equation}
and ${\bf Cov}({\bf X}_s)$ is the covariance matrix of ${\bf X}_s$ \citep{Guy+:2007}.

%----------------------------------------------------------------------------
\subsection{current constraints}
\label{ssec:constraints}
%----------------------------------------------------------------------------
Now we find the best-fitting parameters with the maximum likelihood,
\begin{equation}
-2 \ln {\mathcal L} = \sum_s
\frac{\left[ {\bf V}^{\rm T} {\bf X}_s - M - 5\log_{10}[d_L(\bs{p},z_s)]+5 \right]^2}
{{\bf V}^{\rm T} {\bf Cov}({\bf X}_s){\bf V} + \sigma^2_{\rm int}},
\end{equation}
where $\bs{p}$ represents a cosmological parameter vector, and $z_s$ is SN redshift. The additional variance $\sigma^2_{\rm int}$ accounts for all the sources of diversity of SNe beyond the correction of shape and color and we adopt the value of $\sigma_{\rm int}=0.087$ as suggested in \cite{Guy+:2010}.
We use the light curve parameters $m_B, \Gamma$ and $C$ obtained by the \texttt{SiFTO} model for each SN. 
Under the assumption of flat Universe, the parameters we estimate are $\Omega_{m0}$ and  constant equation of state parameter for dark energy $w_0$ together with the nuisance parameters, $\alpha, \beta$ and $M$.

When we estimate parameters, 
we have to correct the Malmquist bias \citep{Malmquist:1936} with respect to 
distance modulus.
The analysis of the bias with SNLS data sets is studied by 
\cite{Perrett+:2010}.
They use Monte Carlo simulations and 
estimate the relation between source redshift at $ 0.33 \leq z \leq 1.17$ and the dispersion of magnitude due to Malmquist bias.
We correct the bias by subtracting the value estimated by this relation at each SN redshift from distance modulus.

%Magnification $\mu_{\rm lens}$ changes SN flux $F$ to $F \times \mu_{\rm lens}$.
%Then apparent magnitude $m = -2.5\log_{10}(F) + const.$ becomes
%\begin{align}
%	\label{eq:lensed_mag}
%	m' & = -2.5\log_{10}(F\mu_{\rm lens}) + const. \nonumber \\
%    & = m -2.5\log_{10}(\mu_{\rm lens}).
%\end{align}
%We correct the effect of gravitational lensing by 
%$\mu^{\rm obs} \rightarrow \mu^{\rm obs} + 2.5\log_{10}(\mu_{\rm lens})$.

%%%%%%%%%%%%%%%%%%%%%%%%%%%%%%%%%%%%%%%%%%%%%%%%%%%%%%%%%%%%%%%%%%%%%
\section{Estimation of magnification}
\label{sec:est_mu}
%%%%%%%%%%%%%%%%%%%%%%%%%%%%%%%%%%%%%%%%%%%%%%%%%%%%%%%%%%%%%%%%%%%%%
%
\begin{figure*}
\begin{tabular}{cc}
  \includegraphics[width=0.5\linewidth]{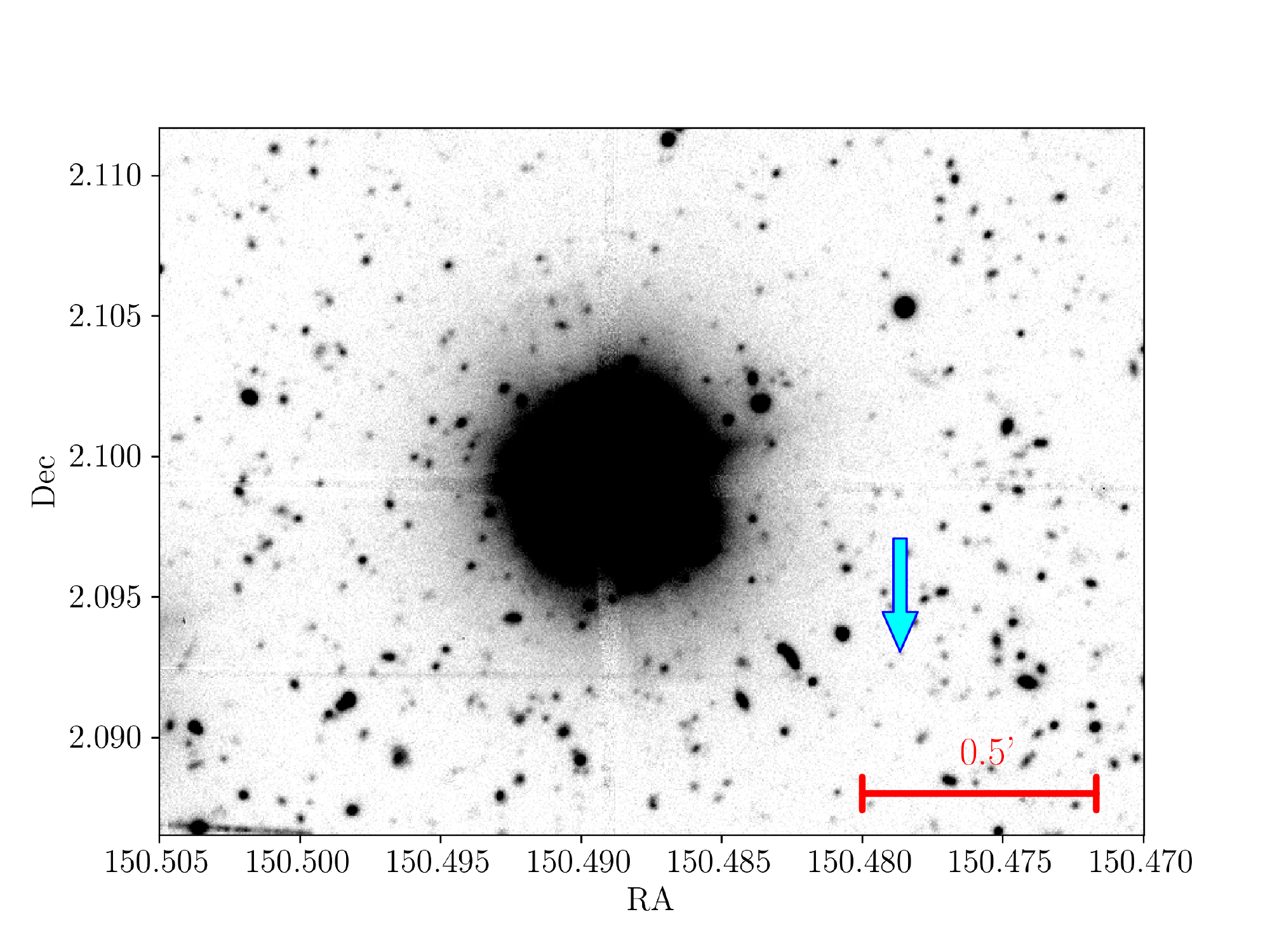}&
  \includegraphics[width=0.5\linewidth]{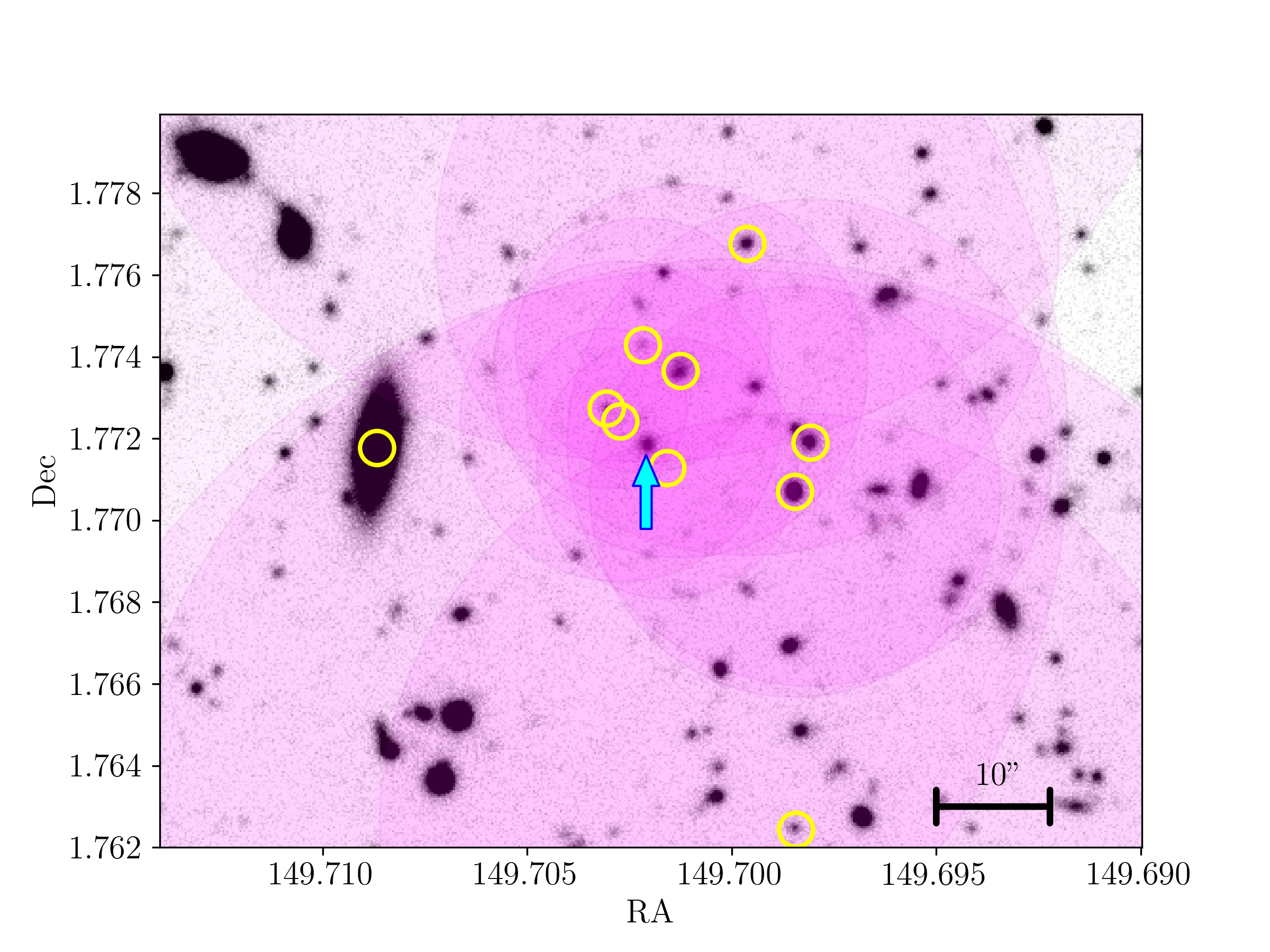}\\
 \end{tabular}
  \caption{
  	(\textit{Left}) An example of SN near bright star.
	Cyan arrow shows the position of SN.
    (\textit{Right}) An example of galaxies we use to calculate magnification.
	Cyan arrow and yellow circles show the position of SN and lens galaxies, respectively.
	Magenta shaded circles are virial radii of lens galaxies.}
    \label{fig:usegalaxy}
\end{figure*}
In this section, we first revisit the estimate of the magnification
which is based on the weak lensing convergence and is widely used in the literature. 
Then we describe our new estimator which may better describe the magnification.
In this paper we consider two different measurements of the magnification for comparison.

%----------------------------------------------------------------------------
\subsection{Convergence measure}
\label{ssec:wl_mu}
%----------------------------------------------------------------------------
We describe the estimator from convergence which is 
reconstructed by the shape of 
galaxies distorted \citep{Oguri+:2018}.
The gravitational lensing distorts the image of the source galaxies.
This effect is described by the Jacobian matrix when the lensed images are projected back to the source plane,
\begin{equation}
	\label{eq:Jacobian}
	{\mathcal A}(\bs{\theta})
	=
	\frac{\partial \bs{\beta}}{\partial \bs{\theta}}
	=
	\left(\delta_{ij}-
	\frac{\partial^2 \psi}{\partial \theta_i \partial \theta_j}
	\right)
	=
		\begin{pmatrix}
		1-\kappa-\gamma_1 & -\gamma_2 \\
		-\gamma_2 & 1-\kappa+\gamma_1
		\end{pmatrix}
		,
\end{equation}
where we define convergence and complex shear $\bs{\gamma}\equiv \gamma_1+i\gamma_2$
in terms of lensing potential $\psi$,
\begin{equation}
	\label{eq:complex_shear}
    \kappa \equiv
    \frac{1}{2} \nabla^2 \psi,\ \ 
	\gamma_1\equiv 
	\frac{1}{2}\left(
	\frac{\partial^2 \psi}{\partial \theta_1^2}
	-\frac{\partial^2 \psi}{\partial \theta_2^2}
	\right),\ \ 
	\gamma_2\equiv
	\frac{\partial^2 \psi}{\partial \theta_1 \partial \theta_2}.
\end{equation}
The magnification can then be
\begin{equation}
	\label{eq:magnification}
	\mu_{\rm lens}
	=
	\frac{1}{{\rm det}{\mathcal A}} 
	=
	\frac{1}{(1-\kappa)^2-|\gamma |^2} .
\end{equation}
Here we work within the weak lensing regime. In the limit of weak lensing where $\kappa, |\gamma| \ll 1$, the magnification can be approximated as
\begin{equation}
	\label{eq:approx_mulens}
	\mu_{\rm lens}
	\approx
	1 + 2\kappa 
	\equiv
	1 + \delta\mu_{\rm lens} .
\end{equation}

The convergence can be related to the shear by
\begin{equation}
	\label{eq:kappafromshear}
	\kappa(\bs{\theta})
	=
	\frac{1}{\pi}
	\int d\bs{\theta}'
	\gamma(\bs{\theta}'){\mathcal D}^*(\bs{\theta}-\bs{\theta}'),
\end{equation}
where ${\mathcal D}$ is a Fourier counterpart of the kernel function,
$\tilde{{\mathcal D}}(\bs{l})=\pi l^{-2}(l_1^2 - l_2^2 + 2i \, l_1 l_2)$
\citep{KaiserSquires:1993}.
Since equation (\ref{eq:kappafromshear}) diverges on small scales, we apply a
two-dimensional Gaussian filter with smoothing scale $\theta_0$,
\begin{equation}
	\label{eq:filter}
	W(\theta)
	\propto
	\exp \left(-\frac{\theta^2}{2\theta_0^2}\right).
\end{equation}

We use HSC shear catalog \citep{Mandelbaum+:2018} with the smoothing scale 
$\theta_0 = 3$ arcmins to reconstruct the convergence maps.
Since the HSC shear catalog only overlaps with \texttt{D1} field, the total number of SNe used for this
convergence measurement is limited to 52. The total number of galaxies used is 104,303.
We use photo-z catalog obtained by {\mizuki} code \citep{Tanaka+:2018}.
For each SN, we reconstruct the surface density using galaxies within SNLS \texttt{D1} field then obtain convergence 
along the lines of sight of SNe from the convergence maps.
In the calculation, we use galaxy whose redshift satisfies
\begin{equation}
\label{eq:photoz_select}
	z_{p, {\rm best}} < z_{SN}.
\end{equation}
After applying this sample selection, we have average galaxy number density $\bar{n}_{\rm gal} = 0.6\, {\rm arcmin}^{-2}$ for $z_s = 0.2$ and $\bar{n}_{\rm gal} = 13\, {\rm arcmin}^{-2}$ for $z_s = 1.0$.

%----------------------------------------------------------------------------
\subsection{Direct measure}
\label{ssec:dist_mu}
%----------------------------------------------------------------------------
%
\subsubsection{Lensing estimation}
\label{sssec:lensing}
Here we propose to measure the magnification in an alternative manner.
In this method, we consider that the SN flux is magnified at the position of foreground galaxies in a single lens approximation.
We also assume
that the galaxy has a spherically symmetric profile, $\rho(r)$.
The projected mass of the galaxy along the line of sight is then,
\begin{equation}
  \label{eq:2dprofile}
  \Sigma(\xi)
  =
  \int \!\! \rho\left(\sqrt{r_z^2+\xi^2}\right) {\rm d}r_z,
\end{equation}
where $r_z$ and $\xi$ are centric comoving radius along and
perpendicular to the line of sight. The convergence and two shear
components induced by the mass associated with the galaxy are then given by
\citep{KaiserSquires:1993},
\begin{align}
  \label{eq:kappa}
  & \kappa(\theta) 
    = 
    \Sigma_{\rm cr}^{-1} \Sigma(D_l \theta), \\
  \label{eq:gamma}
  & \bs{\gamma}(\theta) 
    =
    \frac{1}{\pi}
    \int_{{\mathbb R}^2} \! 
    \bs{{\mathcal D}} (\theta - \theta')\kappa(\theta') {\rm d}^2\theta',
\end{align}
where the kernel function ${\mathcal D}$ is
\begin{equation}
  \label{eq:Dkernel}
  \bs{\mathcal D}(\theta) = \frac{\theta^2_2-\theta_1^2-2i\theta_1\theta_2}{|\theta|^4}.
\end{equation}
The critical surface mass density, $\Sigma_{\rm cr}$, is fully
determined by the distances of lens and source and explicitly given as
\begin{equation}
  \label{eq:sigmacr}
  \Sigma_{\rm cr} 
  =
  \frac{c^2}{4\pi G}
  \frac{D_s}{D_lD_{ls}(1+z_l)^2},
\end{equation}
where $D_{s}, D_{l}$ and $D_{ls}$ are angular diameter
distances from observer to source, lens and from lens to source.
The shear and convergence can be analytically calculated, once assumed 
the mass profile of the galaxy \citep{TakadaJain_kappa:2003, TakadaJain_shear:2003}.
Then the magnification at the sky position separated from center of galaxy by $\theta$ can be calculated as
\begin{equation}
  \label{eq:magnification2}
  \mu_{\rm lens}(\theta) 
  = 
  \frac{1}{[1-\kappa(\theta)]^2-\gamma^2(\theta)}.
\end{equation}

To complete our model, we assume that the galaxy has an NFW
profile given as \citep{NFW:1996}
\begin{equation}
  \rho(r)
  =
  \frac{\rho_s}
  {(r c_{\rm vir}/r_{\rm vir})(1+r c_{\rm vir}/r_{\rm vir})^{2}},
\end{equation}
where $r_{\rm vir}, c_{\rm vir}$ and $\rho_s$ are virial radius,
concentration parameter and overall amplitude.
All those parameters are uniquely determined upon the model calibrated
with the N-body simulation given the virial mass, $M_{\rm vir}$.
The virial radius is often referred as the radius where the total
enclosed mass is equal to the 200 times of critical density of the Universe
and it is related to the virial mass by
\begin{equation}
  r_{\rm vir}
  =
  \left(
    \frac{3M_{200c}}{4\pi\Delta_{200}(z)}
  \right)^{1/3},
\end{equation}
where $\Delta_{200}\simeq 200\rho_{\rm cr}(z)$.
The concentration parameter is related to mass using a suite of N-body
simulation \citep{Duffy+:2008},
\begin{equation}
  c_{\rm 200}(z)
  =
  A(M_{200c}/M_{\rm pivot})^B (1+z)^C,
\end{equation}
where $M_{\rm pivot}=2\times10^{12} h^{-1}M_{\odot}$ and the best fit
parameters for the NFW profile are $A=5.71, B=-0.084$ and $C=-0.47$.
These relations are valid over wide redshift ranges, $0<z<2$ and 
over mass ranges $11<\log (M/M_{\odot}h^{-1}) < 15$.

The halo mass of each galaxy is estimated from the stellar mass obtained from the photometric redshift of HSC. As we described in section \ref{ssec:HSC}, we have two independent stellar mass measurements. We will use both of them to see how much the impact of different measurements of the stellar mass is.
Given the stellar mass of the galaxy, the halo mass can be derived from the stellar to halo mass relation \citep{Behroozi+:2010}.
In order to consider the photo-z uncertainty, the critical surface mass density is weighted by the photo-z probability function as \citep{Mandelbaum+:2008}
\begin{equation}
\label{eq:rachel}
  \Sigma_{\rm cr}^{-1} \rightarrow
  \left\langle
    \Sigma_{\rm cr}^{-1}
  \right\rangle
  =
  \frac{\displaystyle \int_0^{z_s} \! P(z_l) \Sigma_{\rm cr}^{-1}(z_l, z_s) {\rm d}z_l}
  {\displaystyle \int \! P(z) {\rm d}z}
\end{equation}

Total amount of magnification can then be evaluated by 
multiplying 
over all the foreground galaxies,
\begin{equation}
	\label{eq:total_magnification}
    \log \mu_{\rm lens}^{\rm tot}
    =
    \sum_i 
    \log \mu_{{\rm lens},i}(\theta_i) + {\mathcal M},
\end{equation}
where $\mu_{{\rm lens},i}$ is the magnification by $i$-th galaxy, and ${\mathcal M}$ is an average magnification of the Universe. The average magnification ${\mathcal M}$ can be determined so that $\langle \log \mu_{\rm lens}^{\rm tot} \rangle=0$. In our analysis, we calculate the magnification with equation (\ref{eq:total_magnification}) for 1000 random line of sights within the entire SNLS3 and HSC overlapped regions for every redshifts from 0.05 to 1.15 with $\Delta z=0.1$ interval.
We note that the eq. \rref{eq:total_magnification} is only correct when the individual magnification is small and deflection can be negligible. 
%We confirmed that the effect of multiple deflections does not affect our final result at all and we can safely use the approximation of eq. (\ref{eq:total_magnification}) \citep[see ][for more detailed discussion]{McCully+2014}.
Using an updated version of the textsc{gravlens} software  \citep{Keeton:2001}, we calculate the effects of using a full multiplane lensing formalism and find that they are small, confirming that we can safely use the approximation of eq. \rref{eq:total_magnification} \citep[see][for more detailed discussion]{McCully+2014}.
%Since we do not consider the underdense region such as voids in our calculation, 
%estimated magnifications become larger than 1.
%However, from the view of flux conservation, the average of magnification 
%should be 1.
%To avoid this problem, as described in \cite{Kronborg+:2010}, 
%we calculate the average of magnifications for 1000 random lines of sight 
%at each SNLS3 survey fields. 
%The source redshifts are 0.05, 0.15, 0.25, $\cdots$, 1.15.
%Then we subtract the difference between the averaged magnification and 1 
%from the estimated magnification at each SN.
%
\subsubsection{Foreground Selection}
\label{sssec:foregroundselection}
Here we describe the method to 
select the foreground galaxies. First we have to remove the host galaxy of the SN. To identify the host galaxy, we introduce the weighted angular separation $\theta_w \equiv \theta / M_i$, where $\theta$ is a geometrical angular separation between SN and candidate galaxy and $M_i$ is the absolute magnitude of the candidate in $i$-band, which is derived from the photometric redshift ({\mizuki}). We anticipate that the larger absolute magnitude galaxy has more chance to host SN. In the vicinity of SN, we identify the galaxy with smallest $\theta_w$ as the host galaxy. We ignore the contribution to the magnification from the identified host galaxy.
%\co{We do not identify the host galaxy by redshift because of photo-z uncertainty.}

%\begin{figure}
%  \includegraphics[width=\linewidth]{figs/mu_average_s17a.pdf}
%  \caption{
%  	The average of magnifications of each source redshifts.
%    Redshifts are from 0.05 to 1.15 with 0.1 intervals. 
%    Circles, triangles and squares show the averages of magnification 
%    in SNLS3 D1, D2 and D3 field, respectively.
%    Red, blue and green points are the averages of magnification 
%    estimated by photo-z and stellar mass from Mizuki code, 
%    and magenta, lime green and cyan points are estimated from DEmP code, respectively.
%    \label{fig:mu_average}}
%\end{figure}

\begin{figure}
  \includegraphics[width=\linewidth]{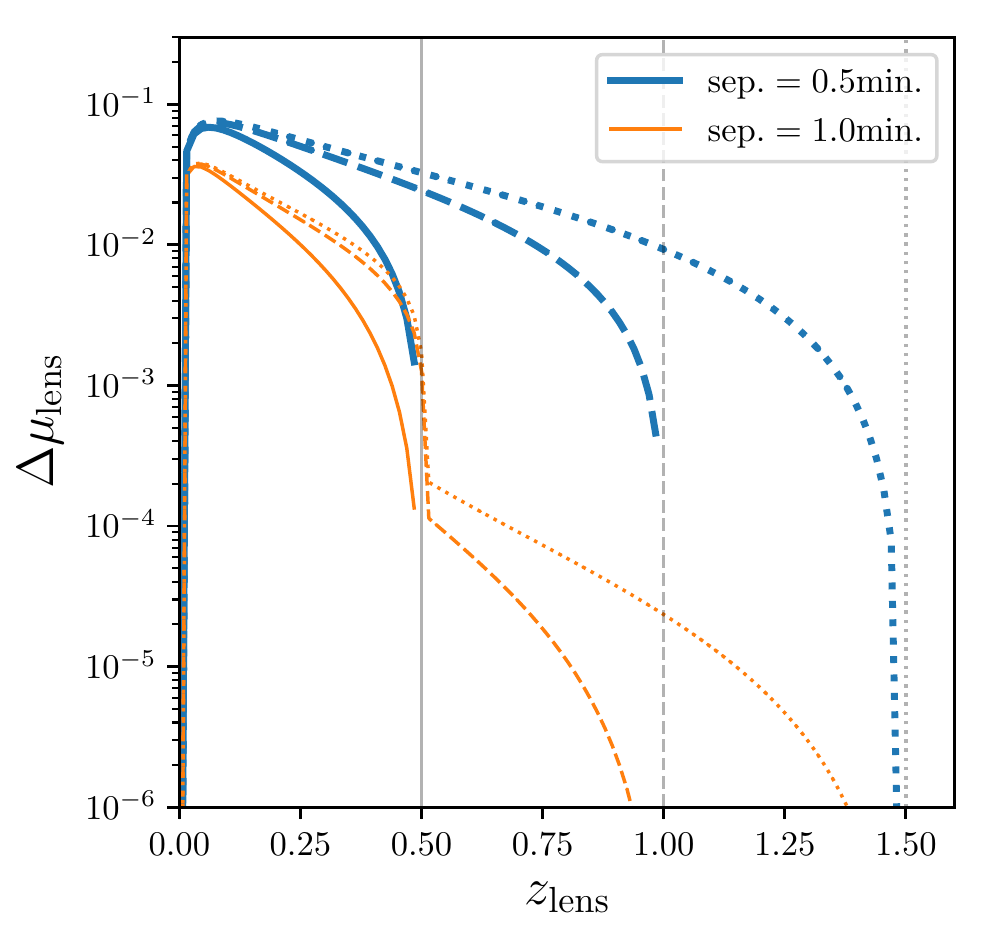}
  \caption{
  Expected contribution to the magnification from $10^{13}\Msun$ halo located at $z_{\rm lens}$ at the separation 0.5 (thick-lines) and 1.0 arcmin (thin-lines) respectively. Solid, dashed and dotted lines correspond to the different source position of $z=0.5, 1.0$ and 1.5. In our model, only shear can contribute to the magnification outside the virial radius, which results significant suppression of $\Delta \mu$ seen at $z>0.5$ for separation$=1.0$ arcmin case.
    \label{fig:zl_mu}}
\end{figure}

Then we select the galaxies which can contribute to the magnification. To select the foreground galaxies, we use all the galaxies where the separation $\theta$ is less than virial radius, i.e. $\theta < r_{\rm vir}/D_l(z_p)$. In practice, background galaxies never contribute to the magnification but it is automatically taken into account by down-weighting by the PDF of photo-z through equation \rref{eq:rachel}.
The example of galaxies we use to calculate magnification is shown 
in the right panel of Figure \ref{fig:usegalaxy}.
Since we assume that the dark matter halo of the galaxy is truncated at $r_{\rm vir}$, convergence vanishes outside $r_{\rm vir}$ but only shear contributes 
to the magnification.
Figure \ref{fig:zl_mu} shows the expected magnification by $10^{13}\Msun$ halo for the SN located at $z_s=0.5, 1.0$ and 1.5. When the separation between SN and lens is 1 arcmin, the magnification drops significantly around $z\sim 0.5$ for $z_s=1$ and 1.5. This is because the separation is larger than virial radii for higher redshift lenses and only the shear contributes to the magnification. As can be seen in Fig. \ref{fig:zl_mu}, the contribution from shear is negligibly small. Therefore, we conclude that the selection of galaxy by $\theta < r_{\rm vir}/D_l$ should be reasonable.

For the galaxy which has its stellar mass larger than 
$10^{11.4}\Msun$, we set the upper limit to the halo mass since in those mass range, the stellar to halo mass relation is not well measured by the simulation. This allows us to avoid too large magnification due to the unreasonably massive galaxy. We set the upper limit as $10^{14.5}\Msun$. 
For our sample, we find only 7.6\% of galaxies for Mizuki and 6.4\% for DEmP exceed this limit.
\subsubsection{Error Estimation}
\label{sssec:error}
Here we describe how to evaluate the error of magnification. The largest sources of uncertainties on the magnification would be photo-z and stellar mass.
For each lens galaxy, 
we randomly draw redshift according to the photo-z PDF measured by \mizuki\, or \demp.
For every nearby galaxies around each SN, the random process may change the foreground galaxy selection, critical mass density of Eq. (\ref{eq:rachel}) but keeping its PDF unchanged. We draw 1000 random samples to evaluate the error.
Together with randomly drawing the redshift,
we also change stellar mass of galaxy 
according to the change on the redshift.
Suppose that stellar mass is proportional to the bolometric luminosity $L$, 
we change stellar mass so that the observed flux $F=L/4\pi D_L^2$ makes unchanged.  
Then the corresponding stellar mass is uniquely determined, once the random redshift is given,
$M_*^{\rm random}=M_*^{\rm best} [D_L(z_{\rm random})/D_L(z_{\rm best})]^2$.
%\ajnc{Any notes on comparison with totally random $M_*?$}

%%%%%%%%%%%%%%%%%%%%%%%%%%%%%%%%%%%%%%%%%%%%%%%%%%%%%%%%%%%%%%%%%%%%%
\section{magnification and Hubble residual correlation}
\label{sec:result}
%%%%%%%%%%%%%%%%%%%%%%%%%%%%%%%%%%%%%%%%%%%%%%%%%%%%%%%%%%%%%%%%%%%%%
%
In this section, we describe the results on the correlation between magnification measured in two distinct methods described in section \ref{sec:est_mu} and the Hubble residual. Then we discuss the effect of magnification on the measurement of cosmological parameters.
%
%----------------------------------------------------------------------------
\subsection{correlation with convergence}
\label{ssec:result_kappa}
%----------------------------------------------------------------------------
\begin{figure}
	\includegraphics[width=\linewidth]{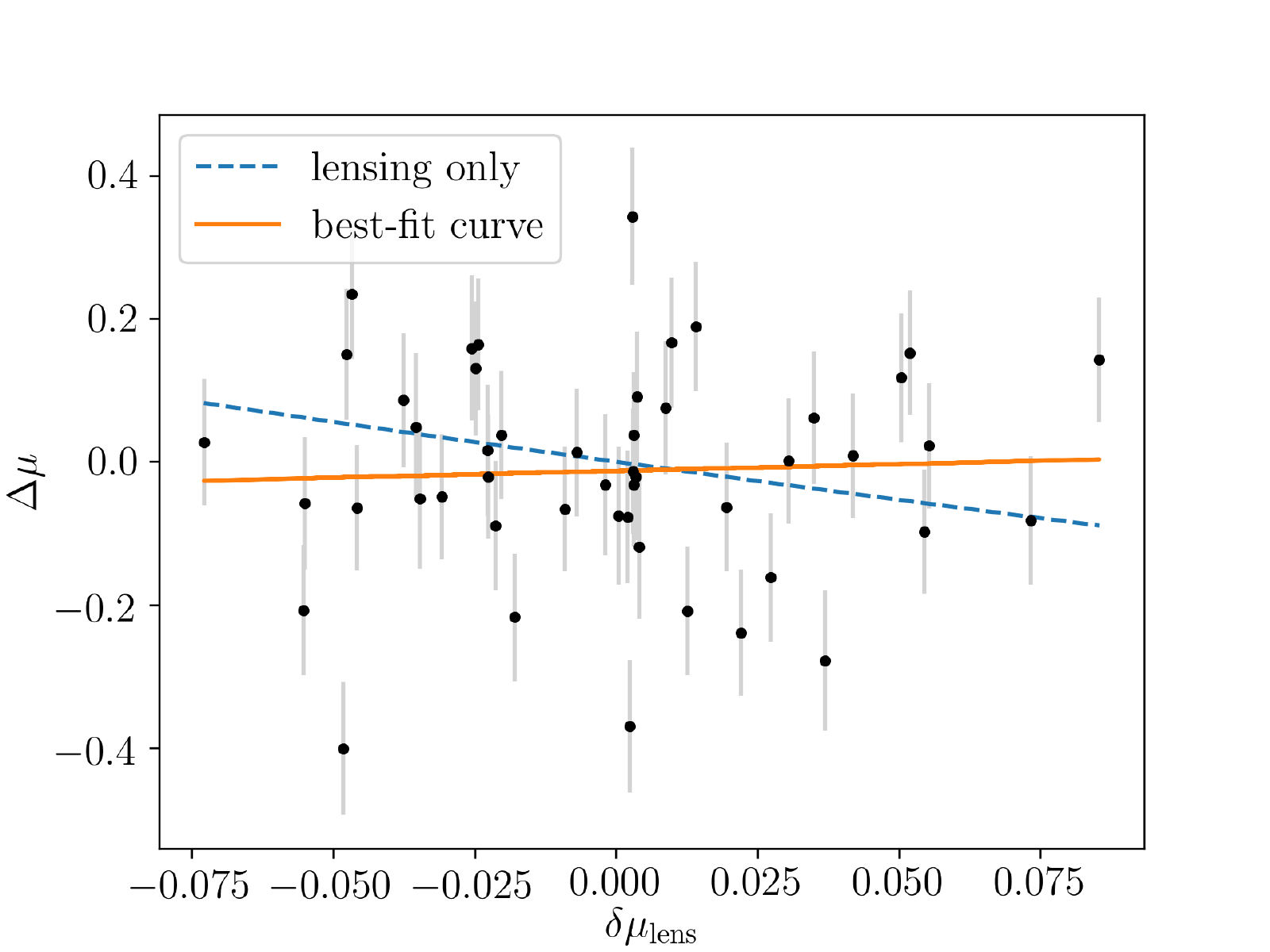}
  \caption{Hubble residual versus magnification obtained by mass reconstruction.
  	We carry out 3$\sigma$ clipping on both convergence and Hubble residual, 
    which removes 3 SNe. Solid and dashed lines are best fitting linear function and expectation from magnification. See the text for more details.
    }
    \label{fig:k_dm}
\end{figure}

\begin{figure}
  \includegraphics[width=\linewidth]{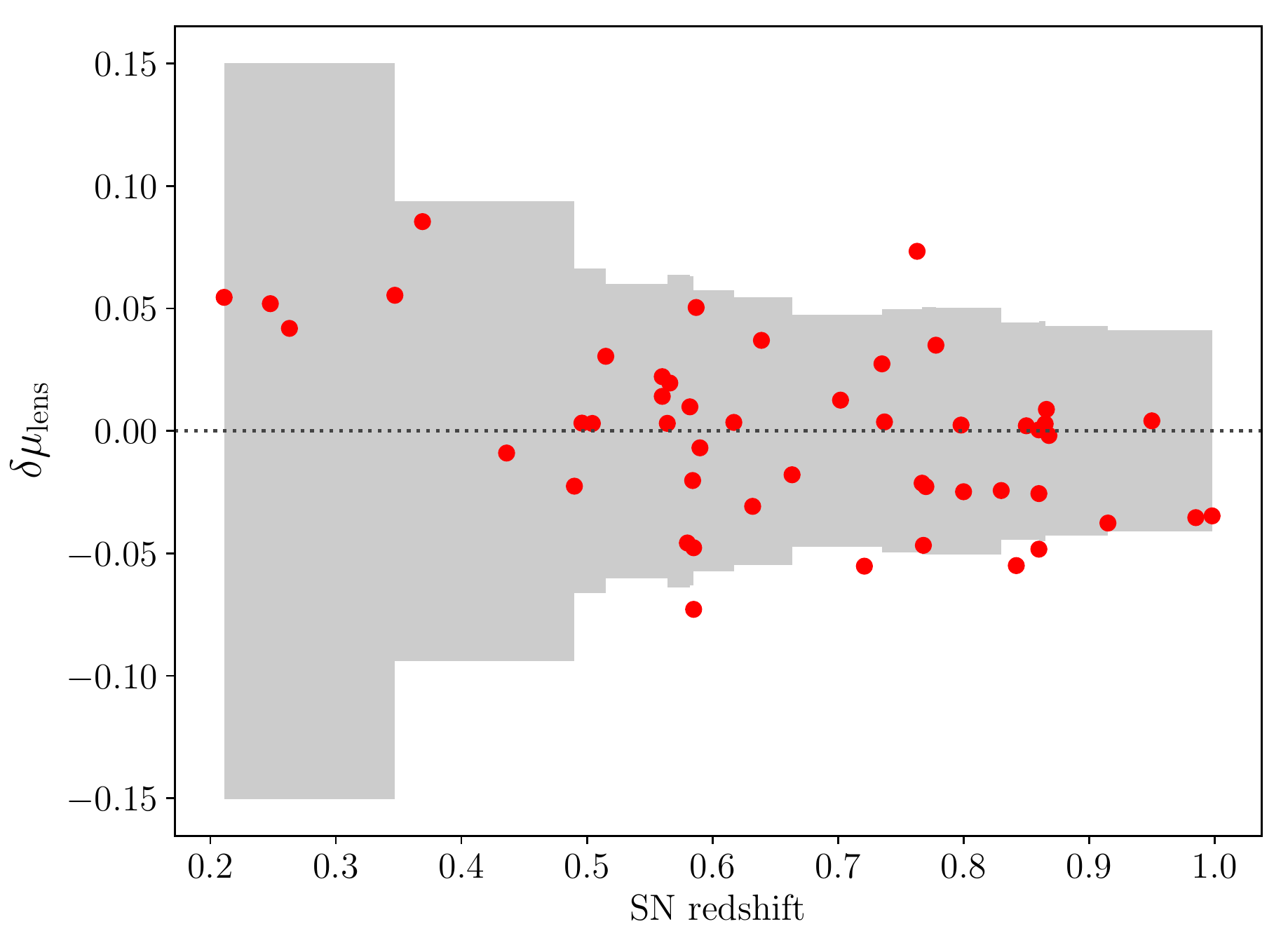}
  \caption{
    Magnification of SNe derived from convergence mass reconstruction. Points are measurement of $\delta \mu_{\rm lens}$ and shaded region is a $1\sigma$ uncertainty derived from 100 random realizations of randomly rotated galaxy ellipticity, averaged over every three SNe positions.
    \label{fig:z_kappa}}
\end{figure}
As we describe in section \ref{ssec:wl_mu}, 
we estimate convergence under the weak lensing approximation and 
search the correlation between convergence and Hubble residual.
Figure \ref{fig:k_dm} 
is a scatter plot for 
Hubble residual 
$\Delta\mu = \mu^{\rm obs} - \mu^{\rm \Lambda CDM}$ of SNe Ia 
and
the magnification $\delta\mu_{\rm lens} = 2\kappa$ at the position of the SN 
reconstructed from the surface mass density using HSC shear catalog.
The orange solid line shows the best-fitting 
linear function and the blue dashed line shows the curve when 
the Hubble residual is perfectly explained by the magnification.
In order to mitigate the effect of outliers on the fit, we
carry out 3$\sigma$ clipping on both convergence and Hubble residual, 
then the final sample shown here is 49 SNe.
In the weak lensing approximation, if the scatters of Hubble residuals are only 
due to the magnification, 
then the $\Delta\mu - \delta\mu_{\rm lens}$ relation becomes 
\begin{equation}
	\label{eq:deltamu_expected}
	\Delta \mu 
	= 
	-2.5\log_{10}(1+\delta\mu_{\rm lens}) \approx -1.086\, \delta\mu_{\rm lens}.
\end{equation}
The best-fitting line we obtain is 
$\Delta \mu = (0.187 \pm 0.364)\delta\mu_{\rm lens} + (-0.013 \pm 0.013)$,
which is consistent with no correlation between the Hubble residual and convergence based magnification.

We further see the correlation coefficient, 
\begin{equation}
	\label{eq:coefficient}
    r =
    \frac{\sum_s (\Delta \mu_s-\overline{\Delta \mu})(\delta\mu_{{\rm lens},s}-\overline{\delta\mu}_{\rm lens})}%
    {\sqrt{\sum_s (\Delta \mu_s-\overline{\Delta \mu})^2}
    \sqrt{\sum_s(\delta\mu_{{\rm lens},s}-\overline{\delta\mu}_{\rm lens})^2}},
\end{equation}
and we find $r=0.032 \pm 0.144$, where the standard deviation of $r$ is obtained by $\sigma_r = (1 - r^2)/\sqrt{n-1}$. It is known that given the sampling correlation coefficient $r$, test of the no correlation for the parent correlation coefficient $\rho=0$ can be done by calculating $t=r/\sqrt{(1-r^2)/(n-2)}$, where $t$ obayes t-distribution with $n$ being the number of samples. In our case, $t=0.22$ and the no correlation, i.e. $\rho=0$ cannot be rejected.

The reason of the no correlation can be fully explained by the noisy measurement of the convergence. To see the measurement accuracy, we compare our results with the random convergence map. The random convergence map is constructed so that the orientation of galaxy is randomly rotated. Figure \ref{fig:z_kappa} shows the magnification signal and random magnification from 100 realizations. It is clearly seen that the magnification from real galaxy ellipticity is well below the random magnification, which means the signal is dominated by the shape noise. Therefore, we do not use the magnification measured by the convergence to correct the scatter of the Hubble residual in the later analysis.
Here we do not use {\demp}, but we expect that the difference of convergence due to the different photo-z code might be small compared to the shape noise \citep[see also ][for photo-z systematic test]{Hikage:2018}. 

Another reason of no correlation is that the smoothed convergence field is not necessarily trace the correct convergence. 
A sufficient number density of background galaxies is required to obtain arcmin scale shear map to mitigate the shot noise and otherwise, it only gives limited value in correcting lensing dispersion of SN Ia \citep{Dalal+:2003}.
It is also shown that the higher order moments such as flexion can reduce the lensing-induced distance errors about 50\% if the galaxy number density is 500-1000 arcmin$^{-2}$ \citep{Shapiro+:2010,Hilbert+:2011}. The average number density of HSC S16A shape catalog is 21.8 arcmin$^{-2}$ \citep{Mandelbaum+:2018} and thus does not suffice for those analysis but will be worth trying for the next generation weak lensing surveys.

%----------------------------------------------------------------------------
\subsection{correlation with galaxy distribution}
\label{ssec:result_mu}
\begin{figure}
  \includegraphics[width=\linewidth]{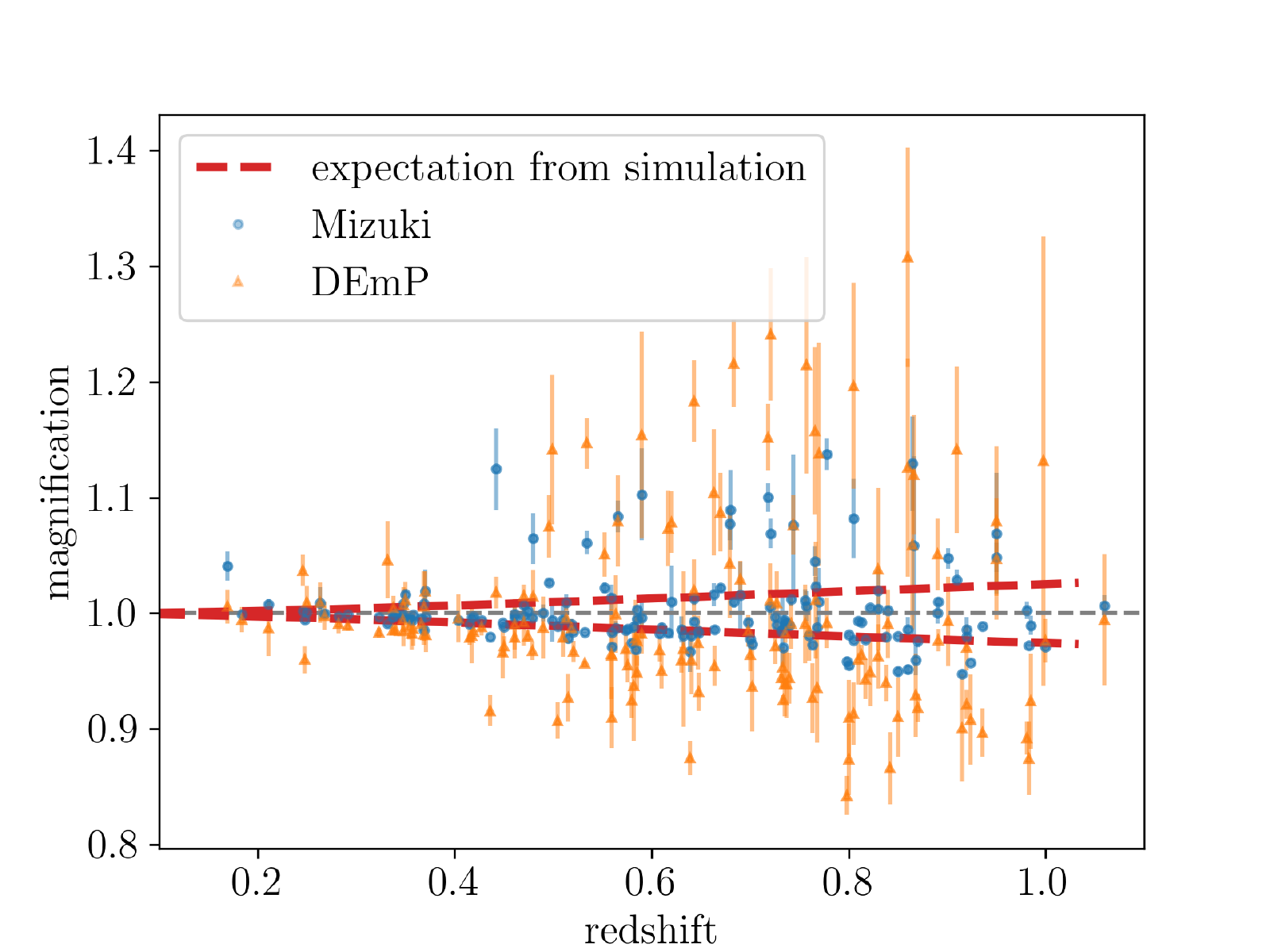}
  \caption{
  	Magnification $\mu_{\rm lens}$ versus redshift of supernovae.
	The blue circles and orange triangles are for {\mizuki} and {\demp}, respectively.
    The red dashed line shows the variance of magnification expected by ray-tracing simulation.
	}
    \label{fig:z_mu}
\end{figure}

%----------------------------------------------------------------------------
As we described in section \ref{ssec:dist_mu},  we estimate the magnification by galaxies along the line of sight, 
assuming an NFW profile for the density profile of dark matter halo.
Figure \ref{fig:z_mu} shows the magnification $\mu_{\rm lens}$ for each SN.
The error bars are calculated by the method described in Section \ref{sssec:error}.
In addition to the foreground galaxy selection in Section \ref{sssec:foregroundselection}, we perform 3$\sigma$ clipping on the magnification to remove 2 outlier SNe when we fit the linear relation, i.e. 151 SNe.
%In this figure, 25 SNe out of 151 are zero consistent with respect to 
%$\delta\mu_{\rm lens} = \mu_{\rm lens}-1$ obtained by Mizuki photo-z and 
%28 SNe are zero consistent for $\delta\mu_{\rm lens}$ obtained by DEmP photo-z.
As can be seen in the Figure \ref{fig:z_mu}, the dispersion of the magnification for {\demp} is larger than that for {\mizuki}. This is due to the difference in the stellar mass measurement between two codes: {\demp} tends to have larger number of galaxies for $M_* > 10^{11} \Msun$.
The more massive galaxy magnifies SN flux more strongly, which causes larger dispersion.
Also, it has larger virial radius and contributes to magnification along multiple lines of sight of SNe.

\begin{figure}
  \includegraphics[width=\linewidth]{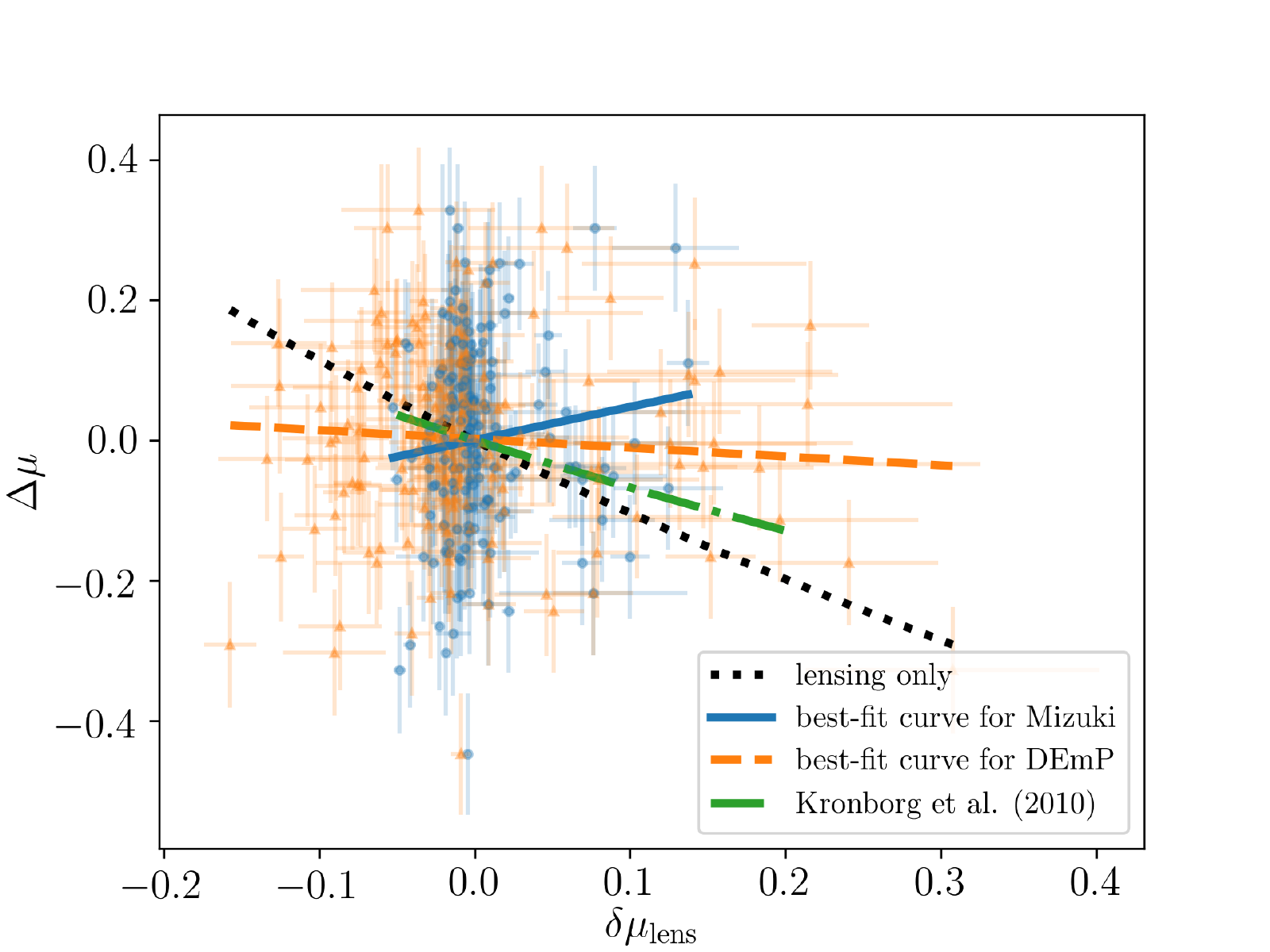}
  \caption{
  Scatter plot for Hubble residual $\Delta\mu = \mu^{\rm obs} - \mu^{\rm \Lambda CDM}$ and $\delta\mu_{\rm lens}$ for 151 SNe.
	The blue solid and orange dashed lines show the fitting curve for {\mizuki} and {\demp},  respectively.
    The black dotted line shows the curve when the scatter is completely described by gravitational lensing effect.
    The green dashed-dotted line shows the best-fit curve obtained by \protect\cite{Kronborg+:2010}.
	}
    \label{fig:mu_dm}
\end{figure}

Figure \ref{fig:mu_dm} plots magnifications $\delta\mu_{\rm lens}$ and 
Hubble residuals $\Delta\mu$, assuming the cosmology summarized in Table \ref{tab:params_fiducial}.
The black dotted line shows the curve when the Hubble residual is completely described by gravitational lensing magnification, 
$\Delta\mu = -2.5\log_{10}(1 + \delta\mu_{\rm lens}) \approx -1.086\, \delta\mu_{\rm lens}$, 
while blue solid and orange dashed line show best-fit curve of our sample 
estimated by photo-z from {\mizuki} and {\demp} code, respectively.
In order to take the uncertainty on magnification into consideration for the fitting, we apply an orthogonal distance regression (ODR) method and find that
%The best-fitting curves are,
\begin{align}
%&\Delta\mu = (0.450 \pm 0.218)\delta\mu_{\rm lens} + (-0.000 \pm 0.007) 
&\Delta\mu = (0.473 \pm 0.221)\delta\mu_{\rm lens} + (0.000 \pm 0.007)
\ \ {\rm for} \ \ {\mizuki}  \nonumber \\
%& \Delta\mu = (-0.109 \pm 0.094)\delta\mu_{\rm lens} + (0.002 \pm 0.007) 
&\Delta\mu = (-0.125 \pm 0.095)\delta\mu_{\rm lens} + (0.002 \pm 0.007)
\ \ {\rm for} \ \  {\demp}.
\end{align}

We calculate the correlation coefficient described equation \rref{eq:coefficient} 
to investigate correlation in Figure \ref{fig:mu_dm} and
find positive correlation $r = 0.070 \pm 0.081$ for {\mizuki} and 
negative correlation $r = -0.037 \pm 0.082$ for {\demp}. 
We test $r$ by the same method described in section \ref{ssec:result_kappa} and find that $t = 0.85$ for {\mizuki} and $t = -0.45$ for {\demp}.
Both estimators cannot reject the no correlation.
\cite{Kronborg+:2010} use 171 SNLS3 SNe to estimate the correlation between Hubble residual 
and the change of distance modulus due to magnification, i.e. $\Delta = -2.5\log_{10}\mu_{\rm lens}$, and find $r = 0.18$, while
\cite{Smith+:2014}, who estimate convergence under the assumption of weak lensing using 608 SDSS SNe, find $r = -0.068 \pm 0.041$.
The correlation coefficient for {\demp} is consistent with the result obtained by \cite{Smith+:2014}.
We also investigate the correlation between $\Delta$ and Hubble residual and find $r = -0.072 \pm 0.081$ for {\mizuki} and $r = 0.026 \pm 0.082$ for {\demp}, which are slightly inconsistent with the result obtained by \cite{Kronborg+:2010}.

%----------------------------------------------------------------------------
\subsection{Estimation of cosmological parameter}
\label{ssec:result_params}
%----------------------------------------------------------------------------

\begin{figure*}
	\includegraphics[width=\linewidth]{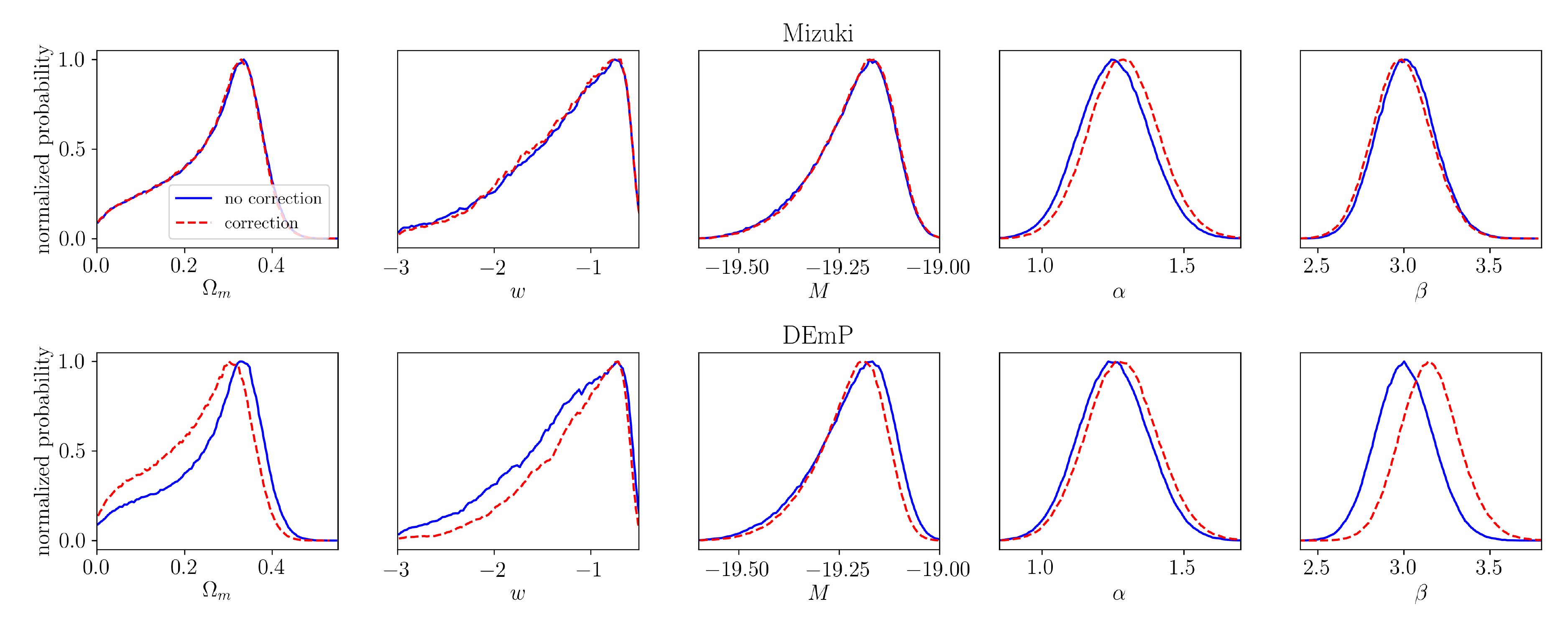}
  \caption{
  Arbitrary normalized posterior distribution function for the parameter marginalized over the other parameters. 
  Blue solid lines are derived by the SNe data set with no correction for the magnification, while red dashed lines are the ones corrected. Upper panels are correction based on the {\mizuki} catalog and bottom panels for {\demp}.
  }
    \label{fig:params}
\end{figure*}

\begin{figure*}
	\includegraphics[width=\linewidth]{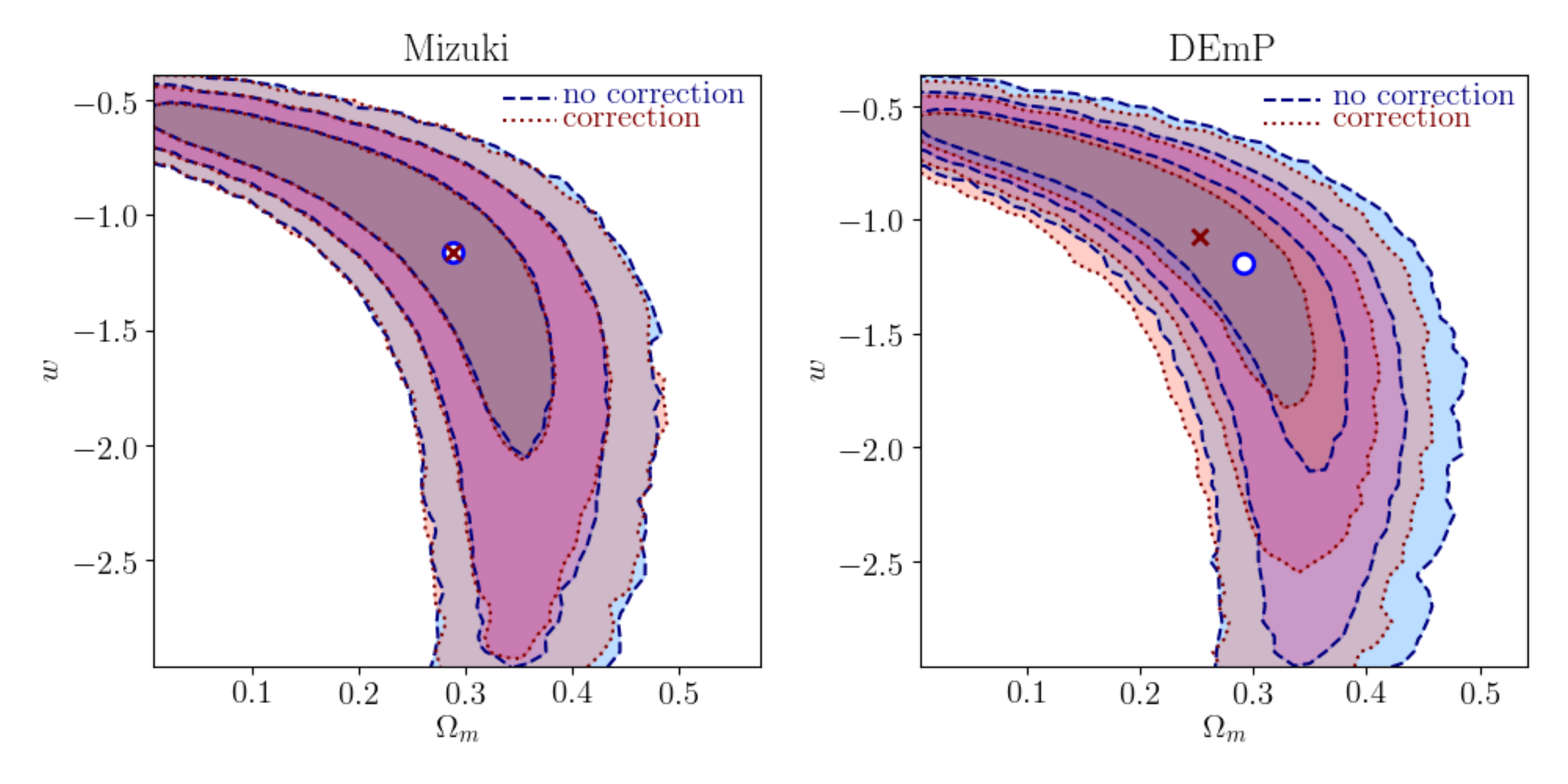}
  \caption{
  Constraints on $\om$ and $w$ from SNe sample corrected for magnification (red dotted) and not corrected (blue dashed). From inner to outer lines, they are 
  68.3\%, 95.5\% and 99.7\% confidence regions of $\om$ and $w$, respectively.
  The blue circles and red crosses show the best-fit values for no correction and correction.
  }
    \label{fig:contour_Om_w}
\end{figure*}

Now we will see the effect of the magnification on the measurement of the cosmological parameters. Since the magnification measurement from convergence is quite noisy, we correct for the magnification derived from the galaxy distribution described in section \ref{ssec:dist_mu}. As in the usual regression, we simultaneously estimate absolute magnitude $M$ and other correction parameters $\alpha$ and $\beta$ together with the cosmological parameters of interest, $\om$ and $w$, which is exactly same procedure with the previous work \citep{Guy+:2010} but limited our sample to SNe overlapped with HSC footprint.
We run the MCMC with Metropolis-Hastings algorithm to get the full posterior distribution function.
Figure \ref{fig:params} shows marginalized 1-dimensional posterior distribution functions.
%with best-fitting parameters in the inset. 
We define the best-fitting value as the median of the marginalized posterior function. The best-fitting values are also summarized in Table \ref{tab:params}.
The difference in the parameters without correction between {\mizuki} and {\demp} can be mainly explained by the different sample selection when we clip out the $3\sigma $ outliers.
We also show two dimensional constraints on $\om$ and $w$ in Figure \ref{fig:contour_Om_w}. If we use {\mizuki}, we find the best-fitting values of $\om$ and $w$ does not change before and after correction. The errors on those parameters are also unchanged. On the other side, the best-fitting parameters when we use {\demp} for the correction, differs slightly: we find slightly smaller $\om$ and larger $w$ after correction but they are still consistent within the 1-$\sigma$ statistical errors. 
Therefore, we find that the photo-z uncertainty does not have much impact on cosmological parameter estimation.
Despite we expected the errors on the cosmological parameters smaller after correction because the magnification correction can reduce the scatter around the theoretical curve, we observe that the error on $\Omega_m$ becomes slightly larger for \demp. On the other hand, the errors on $w$ gets smaller as expected.

\cite{Sullivan+:2011} carry out a joint analysis of SNLS3 with BAO from SDSS and CMB from 7 year WMAP, and obtain $\om = 0.269 \pm 0.015$ and $w = −1.061 ^{+0.069}_{-0.068}$ for the flat Universe model, using 472 SNe.
\cite{Scolnic+:2018} use 1048 SNe Ia sample from Pan-STARRS1 Medium Deep Survey, SDSS, SNLS and Hubble Space Telescope (HST).
They find $\om = 0.307 \pm 0.012$ and $w = −1.026 \pm 0.041$ when combining with Planck2015 CMB results.
The reasons why we obtain worse constraints are (1) the number of our SNe sample is more than three times smaller than those of their samples, and (2) we do not conduct joint analysis  with other experiments such as CMB or BAO.
%\co{When we use {\demp}, we obtain smaller $\om$ and larger $w$ after correction, which lead slightly larger distance modulus for wCDM model at high redshift.
%It is because large magnification at high-z shifts distance modulus too darker.}
%\ajnc{Please write comparison with the previous works, focusing on e.g. the reason why it get worse constraints and/or from other aspects. Also it would be useful to consider why the $\mu$ corrected constraint for DEmP gets smaller value of $\Omega_m$.}

\begin{table}
%\begin{minipage}[t]{0.35\linewidth}
\begin{center}
	\renewcommand{\arraystretch}{1.5}
	\begin{tabular}{crr} \hline \hline
    	 & convergence & direct measure  \\ \hline
         ref.& 
         	Sec. \ref{ssec:wl_mu} & Sec.\ref{ssec:dist_mu} \\
         HSC &
         	HSC-Wide & HSC-Wide/Deep/U-Deep\\
         SNLS &
         	\texttt{D1} & \texttt{D1, D2, D3} \\
         $N_{\rm SNe}$ & 
         	49 & 151 \\
         $\om$ & 
            $  0.198^{+0.090}_{-0.059}$ & $0.253^{+0.050}_{-0.042}$ \\
         $M$ & 
            $-19.210^{+0.074}_{-0.068}$ & $-19.177 \pm 0.040$ \\
         $\alpha$ & 
            $  1.369^{+0.253}_{-0.239}$ & $1.254^{+0.122}_{-0.119}$ \\
         $\beta$ & 
            $  3.809^{+0.509}_{-0.337}$ & $3.029^{+0.171}_{-0.164}$ \\ \hline
    \end{tabular}
    \caption{
   The number of SNe used and fiducial parameters to derive Hubble residual.
   We assume the flat Universe.
	}
    \label{tab:params_fiducial}
\end{center}
\end{table}
\hfill
%
%\begin{minipage}[t]{0.6\linewidth}
\begin{table*}
\begin{center}
	\renewcommand{\arraystretch}{1.5}
	\begin{tabular}{crrrr} \hline \hline
    	 \multicolumn{1}{c}{ }
         	& \multicolumn{2}{c}{\mizuki} 
         	& \multicolumn{2}{c}{\demp} \\ \hline
    	 & uncorrected & corrected & uncorrected & corrected \\ \hline
         $\om$ & 
         	$  0.288^{+0.105}_{-0.086}$ & $0.287^{+0.104}_{-0.085}$ & 
            $  0.292^{+0.102}_{-0.082}$ & $0.253^{+0.113}_{-0.087}$ \\
         $w$ & 
         	$ -1.160^{+0.597}_{-0.363}$ & $-1.161^{+0.595}_{-0.358}$ & 
            $ -1.189^{+0.625}_{-0.354}$ & $-1.078^{+0.498}_{-0.297}$ \\
         $M$ & 
         	$-19.202^{+0.102}_{-0.077}$ & $-19.200^{+0.099}_{-0.077}$ & 
            $-19.205^{+0.102}_{-0.079}$ & $-19.214^{+0.091}_{-0.071}$ \\
         $\alpha$ & 
         	$  1.254^{+0.122}_{-0.125}$ & $1.286^{+0.123}_{-0.123}$ & 
            $  1.252^{+0.120}_{-0.125}$ & $1.279^{+0.124}_{-0.125}$ \\
         $\beta$ & 
         	$  3.011^{+0.161}_{-0.170}$ & $2.990^{+0.165}_{-0.172}$ & 
            $  3.004^{+0.164}_{-0.170}$ & $3.149^{+0.164}_{-0.176}$ \\ \hline
    \end{tabular}
    \caption{
    Best-fit values of each parameters estimated by MCMC method.
    We adopt the median values as the best fit values.
    }
    \label{tab:params}
\end{center}
\end{table*}

\begin{figure*}
	\includegraphics[width=\linewidth]{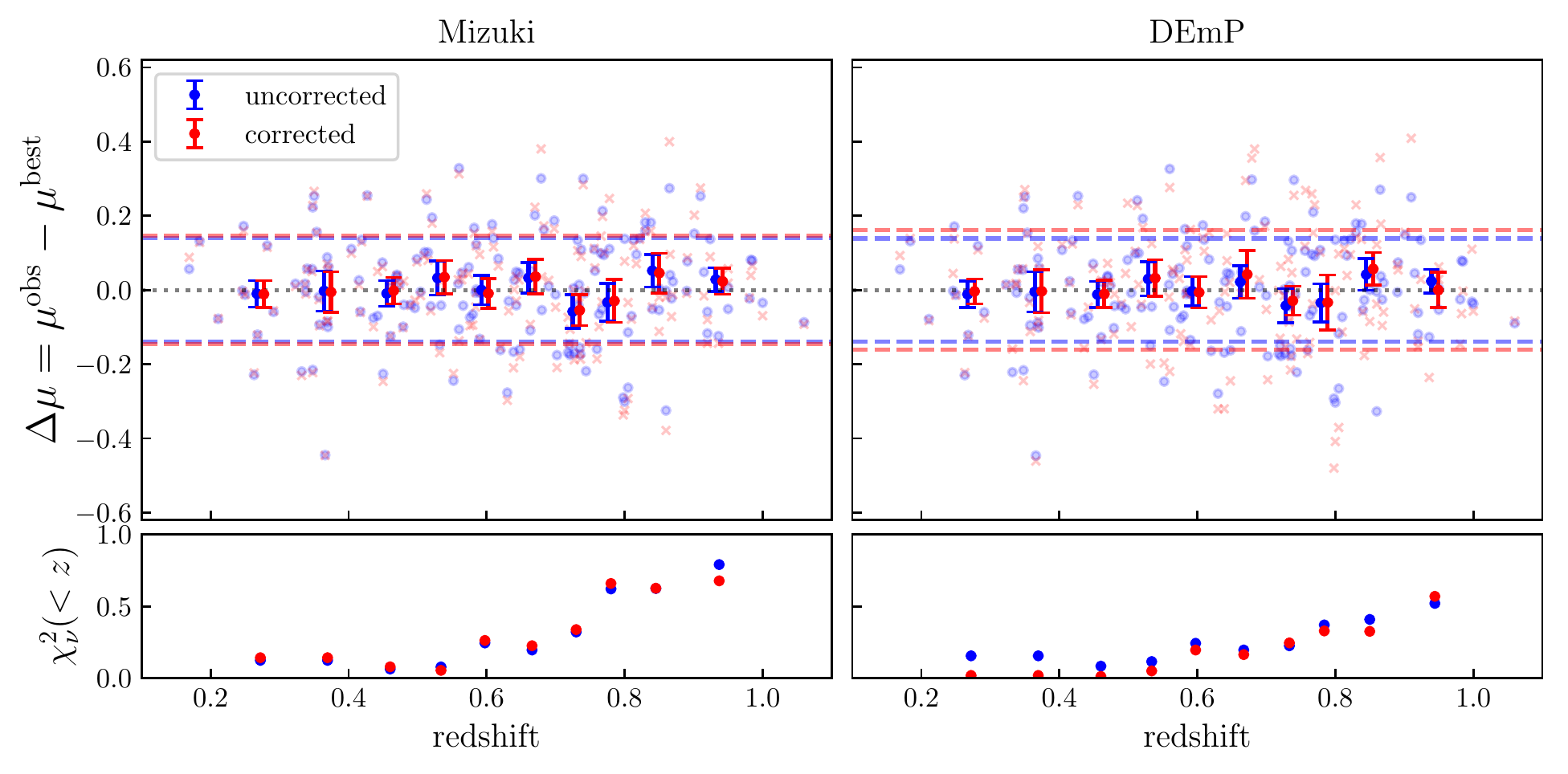}
  \caption{
  (\textit{Upper})
  Hubble residual, $\Delta \mu$ for uncorrected (blue circle) and corrected (red cross) using {\mizuki} (left panel) and {\demp} (right panel). Dashed lines indicate the overall dispersion of $\Delta \mu$, and thick points with errorbars are binned data averaged over every 15 data points. The error bars are $\sqrt{\mean{\Delta \mu^2}/N}$, where $N=15$.
  (\textit{Bottom}) Reduced $\chi^2$ below a certain redshift which is defined in equation \rref{eq:reduced_chisq}.
  }
    \label{fig:residual}
\end{figure*}

Figure \ref{fig:residual} shows how the magnification correction reduces the scatter of the SNe around the best-fitting theoretical curve. In order to quantify the scatter, we calculate the binned reduced $\chi^2$ defined as,
\begin{equation}
\label{eq:reduced_chisq}
	\chi^2_{\nu}(<z)
    =
    \frac{1}{N_{\rm SN}(z_s<z)}
   \sum_{z_a<z} \frac{\mean{\Delta \mu}_a^2}%
   {\mean{\Delta \mu^2}_a},
\end{equation}
where $\mean{\Delta \mu}$ and $\mean{\Delta \mu^2}$ are arithmetic mean and variance within a bin and $N_{\rm SN}$ is the number of SNe below redshift $z$. For the calculation of $\Delta \mu$, we use the corresponding best-fitting model summarized in Table \ref{tab:params}.
While the corrected Hubble residual has slightly larger dispersion as indicated by the dashed lines in the top panel of Figure \ref{fig:residual}, the dispersion averaged over narrow range of redshift is negligibly affected by the magnification correction. As shown in the bottom panels of Figure \ref{fig:residual}, for the case of {\mizuki}, the largest impact of the correction lies on the highest redshift bin which makes dispersion smaller than the uncorrected one. On the other side, for the case of {\demp}, the correction does make dispersion smaller at lower redshifts, but the highest redshift bin contributes to make it larger and this makes overall dispersion slightly larger than uncorrected one. Therefore, we conclude that 
\begin{enumerate}
\item{} the effect of the magnification correction for the SN flux can effectively be ignored and
\item{} the amount of the correction may depend on the photo-z catalog mainly due to the uncertainty on the stellar mass and thus the measurement of the magnification is still not robust.
\end{enumerate}
\cite{Jonsson+:2008} simulated the effect of lensing magnification for SNLS SNe and 
expected that the lensing magnification affects the differences 
$\Delta\om = \om^{\rm lens} - \Omega_m = -0.005$ and 
$\Delta w = w^{\rm lens} - w \approx -0.005$ for 70 SNe.
Our results show that $\Delta\om = -0.001$ and $\Delta w = -0.001$ for {\mizuki} and 
$\Delta\om = -0.039$ and $\Delta w = 0.111$ for {\demp}.
The results for {\mizuki} are consistent with their results, but are not consistent for {\demp}.
\cite{Sarkar+:2008} generated the mock SN samples to estimate the effect of 
gravitational lensing on $w$ and found that the bias on $w$ due to lensing magnification 
is less than 1\%.
%\cite{Smith+:2014} combined their SDSS-II and BOSS SNe samples with WMAP7 CMB power spectrum, SDSS BAO results and SH0ES measurement to constrain cosmological parameters considering the additional term of distance modulus to correct the effect of lensing using number density of galaxy.
%\ajnc{I do not think they use the galaxy number density around SNe to correct for the magnification. If they use it, you should elaborate more on their analysis at section 5.2., what the difference between their method and ours?}
They found that lensing convergences are not affect the central values and 
uncertainties on $\om$ and $w$.

%%%%%%%%%%%%%%%%%%%%%%%%%%%%%%%%%%%%%%%%%%%%%%%%%%%%%%%%%%%%%%%%%%%%%
\section{Summary}
\label{sec:summary}
%%%%%%%%%%%%%%%%%%%%%%%%%%%%%%%%%%%%%%%%%%%%%%%%%%%%%%%%%%%%%%%%%%%%%
In this paper, we have 
applied two distinct 
methods to calculate gravitational 
lensing magnification on supernovae fluxes.
The first method is 
based on the convergence reconstruction,
where we use 49 SNe and $10^5$ galaxies from S16A HSC-Wide shear catalog.
We find that the magnification at the position of the SN has no correlation with the Hubble residual because the local measurement of the weak lensing is quite noisy and convergence signal is fairly consistent with the random.

The second method is directly based on the galaxy distribution around the SNe.
We use 151 SNe and S17A HSC galaxy photo-z catalog to estimate magnification from the projected mass distribution around the SNe.
We use two independent photo-z catalogues, {\mizuki}, a template fitting based catalog and {\demp}, a machine learning based catalog. They both have redshift probability distribution and stellar mass.
We propagate the measurement errors on photo-z and stellar mass to the magnification estimation in a Monte-Carlo manner and find the correlation with the Hubble residual as
$\Delta\mu = (0.473 \pm 0.221)\delta\mu_{\rm lens} + (0.000 \pm 0.007)$ for {\mizuki} and 
$\Delta\mu = (-0.125 \pm 0.095)\delta\mu_{\rm lens} + (0.002 \pm 0.007)$ for {\demp}.
In addition to the linear regression, we see a correlation coefficient and find that 
$r = 0.070 \pm 0.081$ for {\mizuki}, and $r = -0.037 \pm 0.082$ for {\demp}.
This result is consistent with the previous results 
%\citep[][for SNLS3]{Kronborg+:2010} and 
\citep[][for SDSS]{Smith+:2014}.

Finally, we correct the distance modulus of SN for the magnification
to investigate the impact of magnification on estimation of 
cosmological parameters $\om$ and $w$ by MCMC method.
We obtain $\om = 0.287 ^{+0.104} _{-0.085}, w = -1.161 ^{+0.595} _{-0.358}$ for {\mizuki} and $\om = 0.253 ^{+0.113} _{-0.087}, w = -1.078 ^{+0.498} _{-0.297}$ for {\demp},
in comparison with $\om = 0.288^{+0.105}_{-0.086}, w = -1.160^{+0.597}_{-0.363}$ for {\mizuki} and $\om = 0.292^{+0.102}_{-0.082}, w = -1.189^{+0.625}_{-0.354}$ for {\demp} before correction.
We find that they are consistent within 1$\sigma$ errors and magnification has small effect on estimated cosmological parameters.
Our result is consistent with previous results
\citep{Jonsson+:2008,Sarkar+:2008}.

%%%%%%%%%%%%%%%%%%%%%%%%%%%%%%%%%%%%%%%%%%%%%%%%%%%%%%%%%%%%%%%%%%%%%
\section*{Acknowledgments}
%%%%%%%%%%%%%%%%%%%%%%%%%%%%%%%%%%%%%%%%%%%%%%%%%%%%%%%%%%%%%%%%%%%%%
We would like to thank to Ryuichi Takahashi for providing the set of
ray-traced N-body simulations. We also would like to thank to Masahiro
Takada, Nao Suzuki for fruitful discussions.
AN is supported in part by MEXT KAKENHI Grant Number 16H01096.

The Hyper Suprime-Cam (HSC) collaboration includes the astronomical
communities of Japan and Taiwan, and Princeton University.
The HSC instrumentation and software were developed by the National
Astronomical Observatory of Japan (NAOJ), the Kavli Institute for the
Physics and Mathematics of the Universe (Kavli IPMU), the University
of Tokyo, the High Energy Accelerator Research Organization (KEK), the
Academia Sinica Institute for Astronomy and Astrophysics in Taiwan
(ASIAA), and Princeton University.  Funding was contributed by the FIRST 
program from Japanese Cabinet Office, the Ministry of Education, Culture, 
Sports, Science and Technology (MEXT), the Japan Society for the 
Promotion of Science (JSPS),  Japan Science and Technology Agency 
(JST),  the Toray Science  Foundation, NAOJ, Kavli IPMU, KEK, ASIAA,  
and Princeton University.

The Pan-STARRS1 Surveys (PS1) have been made possible through
contributions of the Institute for Astronomy, the University of
Hawaii, the Pan-STARRS Project Office, the Max-Planck Society and its
participating institutes, the Max Planck Institute for Astronomy,
Heidelberg and the Max Planck Institute for Extraterrestrial Physics,
Garching, The Johns Hopkins University, Durham University, the
University of Edinburgh, Queen's University Belfast, the
Harvard-Smithsonian Center for Astrophysics, the Las Cumbres
Observatory Global Telescope Network Incorporated, the National
Central University of Taiwan, the Space Telescope Science Institute,
the National Aeronautics and Space Administration under Grant
No. NNX08AR22G issued through the Planetary Science Division of the
NASA Science Mission Directorate, the National Science Foundation
under Grant No. AST-1238877, the University of Maryland, and Eotvos
Lorand University (ELTE).

This paper makes use of software developed for the Large Synoptic
Survey Telescope. We thank the LSST Project for making their code
available as free software at http://dm.lsst.org.


%\bibliographystyle{mnras}

\bibliographystyle{mnras}
\bibliography{bibdata}

%%%%%%%%%%%%%%%%%%%%%%%%%%%%%%%%%%%%%%%%%%%%%%%%%%

%%%%%%%%%%%%%%%%% APPENDICES %%%%%%%%%%%%%%%%%%%%%
%
%\appendix

%%%%%%%%%%%%%%%%%%%%%%%%%%%%%%%%%%%%%%%%%%%%%%%%%%

% Don't change these lines
\bsp	% typesetting comment
\label{lastpage}
\end{document}